\documentclass[11pt]{article}

\usepackage{amsmath}
\usepackage{amssymb}

\usepackage{graphicx}
\usepackage{epstopdf}

\usepackage{cite}

\usepackage{color} 

\usepackage[labelfont=bf,labelsep=period,justification=raggedright]{caption}

\makeatletter
\renewcommand{\@biblabel}[1]{\quad#1.}
\makeatother

\date{}

\begin{document}

\begin{flushleft}
{\Large
\textbf{Synchronous chaos and broad band gamma rhythm in a minimal multi-layer 
model of primary visual cortex}
}

Demian Battaglia$^{1, 2}$, 
David Hansel$^{3,4\ast}$
\\
\bf{1} Max Planck Institute for Dynamics and Self-Organization, G\"ottingen, Germany
\\
\bf{2} Bernstein Center for Computational Neuroscience, G\"ottingen, Germany
\\
\bf{3} Universit\'e Paris Descartes, Laboratoire de Neurophysique et Physiologie, CNRS UMR 8119, Paris, France
\\
\bf{4} Interdisciplinary Center for Neural Computation, The Hebrew University,
Jerusalem, Israel
\\
$\ast$ E-mail: david.hansel@parisdescartes.fr 
\end{flushleft}

\section*{Abstract}
Visually induced neuronal activity in V1 displays a marked
gamma-band component which is modulated by stimulus properties.
It has been argued that synchronized oscillations contribute to these
gamma-band activity.
However, analysis of  Local Field Potentials (LFPs) across different experiments reveals considerable
diversity in the degree of oscillatory behavior of  this induced  activity.
Contrast-dependent power enhancements can indeed occur over a broad band
in the gamma frequency range and spectral peaks may not arise at all.
Furthermore, even when oscillations are observed, they undergo
temporal decorrelation over very few cycles. This is not easily accounted for
in previous network modeling of gamma oscillations.

We argue here that interactions between cortical layers
can be responsible for this fast decorrelation. We study a
model of a V1 hypercolumn, embedding a simplified description of
the multi-layered structure of the cortex.
When the stimulus contrast is low, the induced activity is only weakly
synchronous and the network resonates transiently without developing collective
oscillations. When the contrast is high, on the other hand,
the induced activity undergoes synchronous oscillations with an irregular
spatiotemporal structure expressing a synchronous chaotic state.
As a consequence the population activity
undergoes fast temporal decorrelation, with concomitant rapid damping
of  the oscillations in LFPs autocorrelograms and peak broadening in
LFPs power spectra.

We show that the strength of  the inter-layer coupling crucially affects this
spatiotemporal structure. We predict that layer VI inactivation should 
induce global changes in the spectral properties of induced LFPs, reflecting 
their slower temporal decorrelation in the absence of inter-layer feedback.
Finally, we argue that the mechanism underlying the emergence of synchronous
chaos in our model is in fact very general. It stems from the fact that gamma
oscillations induced by local delayed inhibition tend to develop chaos when
coupled by sufficiently strong excitation.

\section*{Author Summary}
Visual stimulation elicits neuronal responses in visual cortex. When the contrast of the used stimuli increases, the power of this induced activity is boosted over a broad frequency range (30-100 Hz), called the ``gamma band''. It would be tempting to hypothesize that this phenomenon is due to the emergence of oscillations in which many neurons fire collectively in a rhythmic way. However, previous models trying to explain contrast-related power enhancements using synchronous oscillations failed to reproduce the observed spectra, because they originated unrealistically sharp spectral peaks. The aim of our study is to reconcile synchronous oscillations with broad-band power spectra. 
We argue here that, thanks to the interaction between neuronal populations at different depths in the cortical tissue, the induced oscillatory responses are synchronous, but, at the same time, chaotic. The chaotic nature of the dynamics makes possible to have broad-band power spectra together with synchrony. Our modeling study allows us formulating qualitative experimental predictions, that provide a potential test for our theory. We predict that if the interactions between cortical layers are suppressed, for instance by inactivating neurons in deep layers, the induced responses might become more regular and narrow isolated peaks might develop in their power spectra.


\section*{Introduction}

An increase of activity in the gamma band (30-100 Hz) is 
observed in  Local Field Potential (LFP) and Multi-Unit
Activity (MUA) recordings \cite{Eckhorn1988, Gray1989a, Gray1997, Maldonado2000a, Maldonado2000b,
Fries2001, Logothetis2001, Samonds2004, Henrie2005, Belitski2008, Fries2008, 
Gieselmann2008, Zhou2008, Lima2010}, as well as in EEG and Electrocorticogram 
studies \cite{TallonBaudry1996, Rols2001} in primary visual cortex (V1) upon 
visual stimulation. 
Gamma activity is modulated by properties of the presented 
stimulus, such as orientation \cite{Gray1989a, Lima2010, Kreiter1996}, contrast 
\cite{Logothetis2001, Henrie2005, Ray2010}, velocity \cite{Gray1997, 
Maldonado2000a} or size \cite{Gieselmann2008}, much more strongly than the 
change in power in other frequency bands \cite{Nowak2000, Berens2008}. 
Local GABA-ergic interneuronal networks are thought to play a key role in the production of neuronal activity in the 
gamma range (\cite{Whittington1995}, see \cite{Bartos2007} for a review), 
as upheld as well by recent results obtained through optogenetic techniques 
in-vivo \cite{Cardin2009, Sohal2009}. 

Modeling works have provided a theoretical basis to account for the way in which networks of 
inhibitory interneurons can generate synchronous oscillatory activity in the 
gamma range \cite{Brunel1999, Whittington2000, Brunel2003, Brunel2006, Brunel2008, Wang2010}. 
In brief, in one possible scenario,  the dynamics of the inhibitory post-synaptic currents is 
non-instantaneous (due to axonal delays, but also simply to finite synaptic time-constants). This contributes to create narrow time-windows in which excitatory and inhibitory neurons can fire closely in-phase, before being prevented to do so by a delayed inhibitory feedback. Therefore delayed inhibition, without need of an active involvement of excitatory populations, is capable inducing collective 
synchronous oscillations in neuronal activity. The frequency of 
these oscillations falls in the gamma band if the synaptic time constant of 
the inhibition is in an appropriate range. If a network operates in such a 
synchronous regime the neurons are engaged into 
approximately periodic collective oscillations involving a macroscopically 
large number of neurons. Therefore these oscillations are weakly affected 
by local noise and they maintain coherence  over arbitrarily long time intervals. 
Power spectra of population observables of the network activity 
(e.g. LFP or MUA) exhibit narrow harmonic-like peaks and the 
damping of the corresponding autocorrelograms is slow. 

Peaks in the gamma-band have been identified in the LFP or MUA spectra of 
induced activity in-vivo in V1 \cite{Eckhorn1988, Gray1989a, Gray1997, 
Maldonado2000a, Gieselmann2008}.  However, in general these peaks are very 
broad and in many cases they are virtually indistinguishable as the 
stimulus-modulated gamma power of the signals spreads across a broad-band 
frequency interval \cite{Logothetis2001, Henrie2005, Kayser2004, Nir2007, 
Belitski2008, Briggs2009}. Characterization of the spatio-temporal structure 
of the gamma induced activity by means of auto-correlations (AC) and 
cross-correlations (CC) of single-unit, multi-unit and LFP signals has also revealed
that the neuronal activity has a tendency to oscillate, which can be stronger or weaker, depending on the considered experiment. In some cases the oscillatory components of ACs and
CCs of the induced activity display many cycles before getting damped 
\cite{Eckhorn1988, Gray1989a, Gieselmann2008, Lima2010}. In other cases, however, the oscillations are completely damped after one or two cycles 
\cite{Kreiter1996, Gray1997, Samonds2004, Zhou2008}. The existence of different dynamical regimes might underlie this observed diversity.

For the mathematical abstraction of infinitely large networks, sharp 
boundaries between asynchronous and synchronous dynamical states 
exist \cite{Hansel1996}, but for networks of a finite size such transitions 
are fuzzier \cite{Brunel1999, Hansel1996, Golomb2000}. Consequently, 
if the network does not operate too far from the instability to collective 
oscillations, in a regime which is formally defined as 
asynchronous ---see \cite{Hansel1996, Golomb2000} 
and below for the definition---, the dominant normal modes of the network, 
which describe its response to small perturbations, can display damped 
oscillations at gamma frequencies. Local noise can excite 
these modes, inducing short-lived episodes of synchronous oscillatory activity. 
However, since these episodes are transient, the subsequent
increase in power at gamma frequencies is broad-band.
induced broad band gamma power increases in V1 
can therefore be accounted for if one assumes that the V1 network operates 
in such an asynchronous regime at the edge of developing synchrony
\cite{Rennie2000, Mazzoni2008,Kang2009}. In this regime, correlations in 
the spikes as well as in the membrane potentials of pairs of neurons are 
in general  weak unless the neurons are connected via strong and direct 
synapses. However, in order to get a significant, although damped, oscillatory component in the macroscopic activity, the network must be ``at the edge of synchronization''. Parameters have to be tuned in such a way to be close to an instability toward fully-developed synchronous oscillations, and this tuning have to be tighter, the larger the size of the recruited network \cite{Brunel1999, Brunel2008}. It is not clear how the required fine tuning 
would be satisfied given the range of experimental conditions in which gamma oscillations
have been observed. 

In the present study we explore another scenario which reconciles 
collective synchronous activity with broad-band spectral modulations and robust
fast decoherence. It is based on a mechanism proposed recently for the 
emergence of synchronous chaos 
in recurrent neural networks  \cite{Roxin2005, Roxin2006, Battaglia2007}. 
In this mechanism, clusters of neuron undergo a synchronous gamma oscillation due to local mutual inhibition. These collective gamma oscillations become chaotic when the neuronal clusters are allowed to interact through longer-range excitation.
The resulting overall patterns of activity are characterized 
by synchrony at the population level, but at a same time display a 
characteristic lack of temporal regularity due to  chaos. As a consequence, 
the power of this activity spreads over a broad interval of frequencies 
and the oscillatory components of the autocorrelograms of neuronal activity
and LFP signals are rapidly damped within a few tenths of a millisecond. In this alternative regime, correlations in the spikes of pairs of neurons are still  weak and go together with the sparseness of the firing, but correlations in their membrane potentials can be strong.

We present here a model of a hypercolumn in V1, endowed with a simplified multi-layer 
architecture. In order to explain broad-band contrast-dependent spectral 
modulations in terms of synchronous chaos, we need to identify distinct 
interacting oscillators within the local cortical circuit. We hypothesize that neuronal populations within different thalamo-recipient cortical 
layers are set into oscillation by increased driving and that the mutual 
interaction between these populations, mediated by inter-layer synaptic 
connections, supports the development of synchronous chaos. 
This hypothesis is backed up by anatomical evidence. 
Thalamo-cortical synapses, providing direct sensory-induced driving, 
indeed target cortical layer IV but also, to a lesser extent, 
layer VI  \cite{BlasdelLund1983, Ferster1996, Callaway1998,  Binzegger2004, 
Sincich2005}. Extensive networks of recurrent inhibitory connections are 
present within each thalamo-recipient layer \cite{Binzegger2004, Briggs2010, 
Thomson2010}, supporting local generation of oscillations at multiple 
depths in the cortical tissue. Finally, stereotyped circuit motifs provide 
a bidirectional poly-synaptic connection loop between thalamo-recipient 
layers \cite{Callaway1998, Binzegger2004, Sincich2005, Raizada2003, 
Thomson2003,  Hirsch2006}. 

Relying on extensive numerical simulations, we show that our model displays 
broad-band gamma modulations of the spectra of LFPs upon stimulation
of the network at low as well as at high contrast. Whereas this induced 
activity is asynchronous at low contrast, 
it develops synchrony on a macroscopic scale when the contrast increases.
Therefore we argue that the broad band gamma power observed in recorded
LFP spectra in V1 is compatible with the existence of visually induced
synchronous oscillatory neuronal dynamics.


\section*{Results}

\subsection*{Multi-layer hypercolumn model}

\begin{figure}[!ht]
\begin{center}
\includegraphics[width=4in]{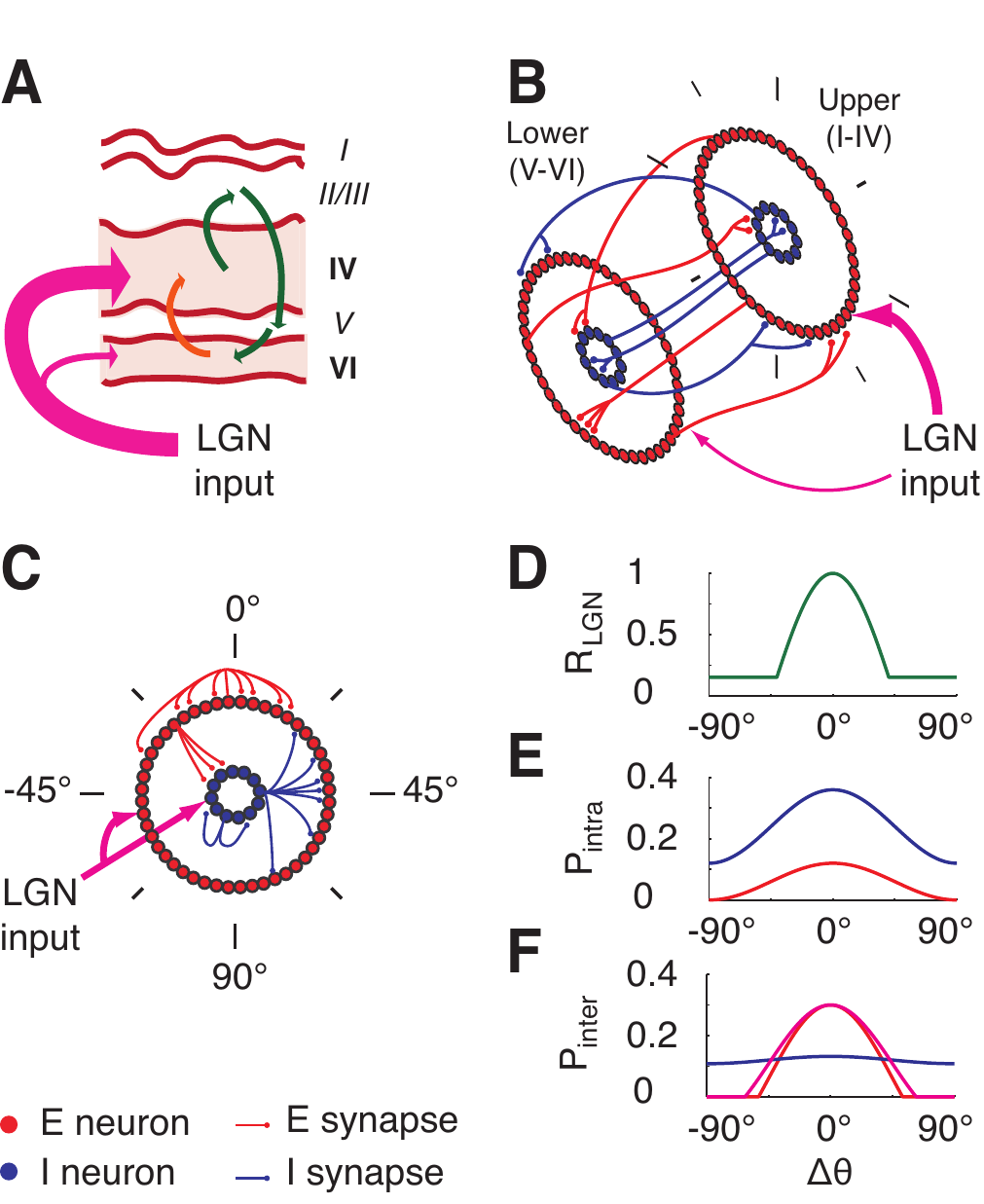}
\end{center}
\caption{}
\label{fig:Circuit}
\end{figure}

We model a functional hypercolumn in primary visual cortex as a large
recurrent network of spiking integrate-and-fire-type neurons.
To account in a simplified way for the
layered structure of the visual cortex ---a cartoon of which is shown
in Figure~\ref{fig:Circuit}A--- the model network consists of two
sub-networks, schematically representing layers I to IV and layers V to VI. 
We denote these two sub-networks as the \textit{upper} 
and \textit{lower layer} respectively (Figure~\ref{fig:Circuit}B).
Each of these layers comprises $N_E$ excitatory
and $N_I$ inhibitory neurons, for a total number of $N=2(N_E+N_I)$ neurons in the network. 
Most of the simulations in this study are performed taking $N_E=4000$ excitatory and $N_I=1000$ inhibitory neurons per 
layer, leading a total of $N=10000$ neurons in the model hypercolumn. This number is one order of magnitude smaller than estimates of the number of neurons in a real V1 hypercolumn based on neuronal densities recently measured by \cite{Stepanyants2009}. However, it leads to dynamical behaviors similar to larger network sizes (see following scaling analyses) and constitutes a compromise for efficient and fast simulations.

Each layer is described by a network with the geometry of a ring as depicted in
Figure~\ref{fig:Circuit}C,
with neurons labeled by angular coordinates, $\vartheta$, ranging 
from -90 to +90 degrees \cite{BenYishai1995, Hansel1998}.
The connections between neurons within each layer
are random, with connection probabilities that depend on the angular
distance between pre- and post-synaptic neurons.
Spatial averages and spatial modulations of connection probabilities are
set independently for the various kinds of connections
(e.g. excitatory-to-excitatory, excitatory-to-inhibitory,
inhibitory-to-excitatory or inhibitory-to-inhibitory), thus making it possible to
vary the spatial profiles of net synaptic interactions
(see Figure~\ref{fig:Circuit}D, E, F). Excitatory 
and inhibitory inter-layer connections are also random and spatially modulated.
All the external inputs to the network are modeled as stochastic processes (see Methods section).
The neurons receive an external non-selective noisy current representing
background inputs to V1 from other cortical and subcortical areas
and a weakly tuned noisy current which represents
visually induced inputs to V1 from converging Lateral Geniculate Nucleus (LGN) synapses \cite{Hubel1962}. 
Note that the two main thalamo-recipient layers, i.e. layers VI and IV, are embedded within two distinct model layers. 

Our two-layer circuit embeds in a simplified manner several
known features of the stereotypical interlaminar anatomy of the columnar
microcircuit, in particular, the existence of a layers IV to VI to IV
feedback loop  \cite{Callaway1998,  Thomson2003, Sincich2005}.
Furthermore,  a different degree of
spatial modulation for inter-layer excitation and inhibition mimic the on-center off-surround arrangement of layers VI to IV
projections \cite{Ahmed1997}.
In the simulations described below  we assume that the
LGN input to the lower layer is weaker (by a factor of 2) than the input to
upper layer to account for the fact that thalamo-cortical
synapses reaching layer VI are smaller in number than those reaching layer IV \cite{Binzegger2004}.
We also assume that latencies for inter-layer connections are longer
than for intra-layer connections, thus accounting for the multisynaptic
nature of this coupling. Our assumptions on the connectivity, external inputs
and latencies  are further commented upon in the \textit{Discussion} section.

In order to analyze the role of the interlayer interactions in shaping the
spatiotemporal dynamics of our model hypercolumn, we introduce a  parameter
$ 0 \leq \Gamma \leq 1$ which homogeneously rescales the strength of excitatory
and inhibitory connections between layers.
For $\Gamma = 1$ the interactions
between the layers assume their maximum strength. For $\Gamma = 0$ the layers
are completely independent. In the following, we consider first the dynamics
of the network at full coupling strength, $\Gamma = 1$.

\subsection*{Orientation tuning and contrast dependence of induced response}

In absence of ``visual'' stimuli (contrast level
$C = 0\%$), the model hypercolumn is driven only by the non-selective
background input. The resulting spontaneous activity is heterogeneous across
the neurons with average firing rates of $1.2 \pm 0.4$ Hz and $5 \pm 3$ Hz for
excitatory and inhibitory neurons, respectively. Differences in the
spontaneous firing rate distributions for upper and lower layers are not
statistically significant at the 5\% confidence level. The spontaneous
firing of the neurons is highly irregular due to the stochasticity of the inputs. For instance, the average
coefficient of variation (CV) of the interspike histogram of excitatory 
and inhibitory neurons in the upper layer is CV$~ = 0.9\pm 0.2$. More details about rate and CV distributions can be found in
Appendix S1.

The profile of the activity induced by an oriented stimulus 
in both layers, is localized and centered at an angular coordinate 
corresponding to the stimulus orientation. Hence, the neuronal 
responses are selective to the stimulus orientation. The tuning curves
of individual neurons display some heterogeneity in their broadness, as
exhibited by distributions of peak response rates, circular
variance and skewness of the tuning curves (reported in Appendix S3).

Figure~\ref{fig:TuningCRF}A displays the population
average tuning curve for various contrast levels 
for excitatory neurons in the upper layer.
Comparison between tuning curves at different contrasts reveals that tuning 
width is approximately contrast invariant and that the larger deviations are
observed for small contrast levels (tuning curves normalized to the peak are 
plotted in Appendix S4).
This invariance is achieved as an effect of noise in synaptic inputs 
\cite{Anderson2000, Hansel2002}. 

The preferred responses
of the excitatory neurons vary non-linearly with the contrast 
as depicted in Figure~\ref{fig:TuningCRF}B, where the population 
average Contrast Response Functions (CRFs) are plotted for excitatory  
neurons in the upper layer.  It can be fitted by an hyperbolic ratio 
function (see {\it Methods} section), 
with mid-range contrast $C_{50} \approx 15$\% and an exponent of $n\approx 5$ 
(upper layer neurons).  This nonlinear dependence stems from the fact that 
increased sensory-driving yields larger inhibitory neurons activity which in turn is responsible for the saturation of the excitatory population 
response \cite{Persi2010}.
The CRFs of inhibitory neurons show a much weaker tendency to saturation 
at large stimulus contrasts which is due to the logarithmic dependency
on the contrast of their external input.  
The CRFs of single neurons are heterogeneous, in qualitative agreement with 
experimental reports \cite{Contreras2003} 
(see Appendix S3). The contrast response functions of 
the lower layer are homologous, but the 
induced responses are approximately twofold smaller, due to the weaker 
LGN driving. 

\subsection*{The dynamical state of the network depends on the stimulus contrast}

For zero contrast, the synchrony level in the spontaneous neuronal activity is small, as denoted by a small value of the synchrony factor $\chi$. This factor, defined in the Methods section, quantifies global synchrony over a network and is bounded between 0 and 1. For a network of size $N=10000$, the synchrony factor for spontaneous activity assumes the value $\chi=0.02$. Furthermore, as shown in Figure~\ref{fig:Chi}B (grey line), it vanishes consistently as $\chi\propto {\frac{1}{\sqrt{N}}}$ for larger network sizes, allowing us to classify formally the (asymptotic) state of the network as ``asynchronous'' (see Methods section and \cite{Hansel1996, Golomb2000}).

The single neuron and population responses of the network
to a low stimulus contrast, $C=2\%$ is illustrated in 
Figure~\ref{fig:LowContrast}.
The raster plot of the spike activity of all the excitatory neurons in the 
upper layer is
plotted in Figure~\ref{fig:LowContrast}A. It suggests that the firing is highly irregular (the mean CV of the upper layer excitatory neurons is $0.9 \pm 0.1$, see Appendix S1) and that the network activity of the 
network is only weakly synchronized.
This is confirmed in Figure~\ref{fig:LowContrast}B where the
spike trains  of six upper layer cells stimulated within $\pm 5^\circ$
from their preferred orientation are plotted. The neurons fire without any noticeable
synchrony. Figure~\ref{fig:LowContrast}C displays the voltage traces of two
of these neurons.
The comparison between the sub-threshold fluctuations in the two traces 
does not reveal any significant correlation.
To further quantify the correlations in the supra and subthreshold activity of the neurons 
we compute the zero delay pairwise correlation coefficients (CCos) of the spikes and the 
membrane potential traces for a large number of pairs formed by highly active neurons with preferred orientation within $\pm 9^\circ$ from the presented stimulus 
(see \textit{Methods} section and Appendix S2 for
details). The resulting histograms  are shown in Figure~\ref{fig:LowContrast}D
(spikes: left, cyan color; voltage: right, blue color). They are peaked around zero
with a mean statistically indistinguishable from zero ($0.000\pm 0.001$ for spikes and voltage). Almost all the CCos are  weak for the spikes as well as for
the voltage traces  (CCos larger than 0.25 occur only for 2\% of the pairs when considering spike CCos, and for 0.1\% of the pairs when considering voltage trace CCos). These results are consistent with a very weak synchronization in the network
activity. This is in line with the  small value of the synchronization factor,
which is only $\chi = 0.03$. Auto- and crosscorrelograms of spike trains and membrane potential traces of three representative neurons are also shown in Figure~\ref{fig:PairwiseCCs}A,B.
The pairwise crosscorrelograms of both spikes and voltages  do not display any persistent oscillatory component, even when two cells share a same orientation preference.

The dynamical state of the network is qualitatively different for a high contrast
stimulus. For  $C=95\%$ the neurons are engaged into a collective pattern of 
synchronous oscillations in contrast to what happens for $C=2\%$. 
This is clear from the raster plot in Figure~\ref{fig:HighContrast}A. 
Figure~\ref{fig:HighContrast}C plots the membrane potential traces of 
two neurons. Comparison of these traces suggests that now the subthreshold 
membrane fluctuations of the neurons are strongly correlated across the 
network. As a matter of fact, the  synchrony factor, $\chi$, which characterizes
the degree of synchrony in the subthreshold activity at the network level,
is $\chi = 0.75$. However action potentials are much less synchronized, as suggested by the comparison of the spike trains of the six
neurons plotted in Figure~\ref{fig:HighContrast}B: although 
multi-neuron coincidences in firing (denoted by vertical grey bars) can be 
detected, the overall  synchrony is weak. 
This substantial difference in the strength of the 
pair correlations in supra and subthreshold activities
is clear in Figure~\ref{fig:HighContrast}D. All the CCos 
of the subthreshold membrane potentials (red histogram) are large and sharply 
distributed around  0.75 (standard deviation of $\pm 0.03$) whereas
the distribution of the spike trains CCos (magenta histogram) has a mean which
is only $0.097\pm 0.002$. Remarkably, the firing activity continues to be highly irregular, despite the high degree of synchrony (mean CV of upper layer excitatory neurons is CV= $1.0\pm 0.2$, see Appendix S1).
 Auto- and crosscorrelograms of spike trains and membrane potential traces of three representative neurons are shown in Figure~\ref{fig:PairwiseCCs}C,D.
The pairwise crosscorrelograms of voltages display now a clear oscillatory structure, which is however completely damped after only two or three cycles. Note that  oscillatory correlations are evident even when the difference of preferred orientation is large ($> 20^\circ$). Note as well that pairwise crosscorrelograms of spike trains do not display any marked oscillation even when the two considered cells have  similar preferred orientations.
We stress that the small mean value CCos and the lack of a clear oscillatory structure in the crosscorrelograms for spike trains, in both the low and the strong contrast case, is associated to the irregularity and the sparseness of single neuron firing. 

These results indicate that  synchrony in the population
activity increases with the contrast. As a matter of fact, 
the synchrony measure $\chi$ varies abruptly around a contrast value
of $\sim$10\%, as shown in Figure~\ref{fig:Chi}A. This
is even sharper with larger network sizes (compare in Figure~\ref{fig:Chi}B,
the solid line which is for $N = 10000$ with the dashed line which is for
$N= 40000$). Moreover, a systematic analysis of the dependency 
of $\chi$ on the size $N$ reveals that for $C \lesssim 10\%$, $\chi$
({\it low contrast regime})vanishes consistently with $N$,  $\chi\propto {\frac{1}{\sqrt{N}}}$,
while for $C \gtrsim 10\%$ ({\it large contrast regime}) it converges toward a constant non zero value 
(Figure~\ref{fig:Chi}B). Hence, the network operates in qualitatively
different regimes at low and high contrast. Whereas the network state
can be classified as asynchronous in the low contrast regime, it is synchronous in the high contrast regime. 
This sharp variation of synchrony is indicative of a phase-transition occurring for increasing contrast, due to an increased drive to the network (see Discussion).

To characterize further how the population dynamics  depend 
on the contrast we compute the autocorrelation, $AC(\tau)$  
of the LFP signals induced by 
stimuli oriented at the preferred orientation of the recording site 
(see \textit{Methods} for the way we relate the LFP signals to
the neuronal activities in the framework of our model and Appendix S6 for examples of LFP traces). The result for 
low contrast, $C=2\%$, is 
plotted in Figure~\ref{fig:LFPAC}A, B. The amplitude of the (non-normalized) AC at zero delay,
$AC(0)$, is small and decreases with the network size as $1/\sqrt{N}$.
Similarly, the small oscillatory component 
of the AC disappears gradually for increasing network sizes 
(Figure~\ref{fig:LFPAC}B).
This is because the network state is asynchronous and 
in a larger network more cells contribute to the LFP signal 
(see also \textit{Methods} section).

The fact that at high contrast, $C=95\%$, the network is engaged
in collective synchronous activity is manifest in Figure~\ref{fig:LFPAC}C,D:
$AC(0)$ is now large and it does not vanish in the large $N$ limit
and is almost independent of $N$ for $N > 10000$.  However, and remarkably, 
the induced dynamics exhibit a spatio-temporal structure which is more complex 
than a periodic regular oscillation of the population activity:
the time interval between consecutive episodes of synchronous activity 
displays cycle-to-cycle fluctuations as can be observed in the
raster plotted in Figure~\ref{fig:HighContrast}A).
As a result, the LFP autocorrelogram is rapidly damped. 
Although it displays some secondary peaks their amplitudes are very small
as shown in Figure~\ref{fig:LFPAC}C. The damping of the AC oscillations
is even faster for larger network sizes (Figure~\ref{fig:LFPAC}D). Note that autocorrelations for intermediate contrast values are also rapidly damped (see Appendix S6). {A moderate tendency to period doubling, manifested by a second autocorrelogram peak slightly larger than the first autocorrelogram peak, has not been reported experimentally. We remark however that this is an accidental feature, which is no more observed, for instance, for larger network sizes or stronger inter-layer coupling.}

LFPs induced by non-preferred stimulus directions display as well oscillatory components, for both low and high contrasts. Induced LFPs are correlated over the entire ring network as revealed by crosscorrelation analysis, confirming that sub-threshold coherence can exist independently from correlations in spiking activity (see Appendix S6).

Finally, {we consider the spectral properties of induced LFPs, and their relation with MUA observed at a same location.} The dependency on the contrast of the power spectra of the LFPs 
induced by preferred-orientation stimuli is shown in Figure~\ref{fig:LFP}{A}.
The low-frequency part of the power spectra is weakly dependent 
on the stimulus contrast. Rather, it is shaped by the properties of 
cortical background activity, modeled as a stochastic Ornstein-Uhlenbeck noise 
with a frequency cutoff (see \textit{Methods} section and  \cite{Rudolph2004}). 
This should be compared to the boosting of the power 
as the contrast increases for frequencies $> 30$ Hz.  
Although the network activity becomes much more synchronous  
at large contrast as explained above, power spectrum modulations 
are not limited to narrow peaks, but, even at the highest contrast, 
the whole frequency range comprised between 30 and 100 Hz is boosted.
{In this same broad frequency range in which contrast-dependent power modulations occur, the LFP displays phase-synchronization with the MUA at a same location, as measured by a MUA-LFP coherence increasing with contrast (see Figure~\ref{fig:LFP}B). Interestingly, the MUA-LFP coherence, even at full contrast, rises only at an average peak level of approximately 0.3, compatible with physiologic ranges of synchronization \cite{Henrie2005, Womelsdorf2007}. This can be explained by the random-like variability of single neuron firing ---inherited by the MUA signal, which reflects the spiking activity of only a limited number of single units (see \textit{Methods} section)---, but also by the lack of phase autocoherence in the LFP signal itself (cfr. \cite{Burns2010}).
}   

The spatio-temporal structure of the induced activities in the lower 
and in the upper layers are similar. In our simulations, the 
lower layer average firing rate is approximately half of that in the 
upper layer, reflecting weaker driving from LGN. Cross-correlation analysis  of the LFPs in the two layers  shows that 
the lower layer oscillations lag behind those in the upper layers (see Appendix S5). Note that  larger response latencies in deep layers have been experimentally observed in specific conditions \cite{Wallace2008, Sakata2009}. However, the multi-layer structure in our model is too schematic to capture quantitatively such inter-layer relations. In particular, the difference in response rate and the exact locking pattern between layers are not robust in our model against changes in the parameters of LGN input and inter-layer coupling. On the contrary, the synchronization and the fast decorrelation of induced oscillations are robust qualitative properties (see later Discussion).

\subsection*{The role of inter-layer coupling in destroying the temporal
coherence of the oscillations}

In order to explore the role played by the inter-layer interactions, we investigate in the following how the dynamics in the high contrast regime is affected by a change of this coupling. More specifically, we rescale the peak conductances of all the synapses between cells in different layers by a same factor $0 <  \Gamma < 1$ (here $\Gamma  = 1$ and   $\Gamma= 0$ correspond respectively to fully coupled and fully decoupled layers).

Upon  layer-decoupling the mean firing rate of the excitatory and inhibitory cells 
increases in the upper layer 
(Figure~\ref{fig:Decoupling}A). However response rate changes are 
highly heterogeneous across cells and, in some cases, the peak rate is even 
slightly reduced. An analogous heterogeneity is observed in the changes of 
preferred orientation, skewness and tuning width. However, even though changes 
after complete layer decoupling can be significant for specific cells,  
the distribution of tuning curve parameters over the entire upper layer 
excitatory neurons population is only weakly altered. Details are shown 
in Appendix S7. 

Another effect of layer decoupling, albeit moderate,
is that the degree of synchrony in induced activity decreases monotonically with 
$\Gamma$ (Figure~\ref{fig:Decoupling}B). 
For instance, the synchrony factor is $\chi=0.75$ for $\Gamma=1$, 
but decreases to $\chi=0.71$ when $\Gamma=0.5$, and drops further to 
$\chi=0.67$ for fully decoupled layers. 

The most striking consequence of the reduction in inter-layer
coupling is the progressive qualitative change in the  shape of 
the LFP autocorrelograms and power spectra as $\Gamma$ decreases. 
This is depicted in Figure~\ref{fig:Decoupling}.
For 80\% coupling strength ($\Gamma = 0.8$), the autocorrelogram of 
LFP and the corresponding power spectrum are similar to what is
found in the fully-coupled case (fast temporal decorrelation and 
broad plateau-like peak in the gamma spectral band, see 
Figure~\ref{fig:Decoupling}C, D). However, for a 60\% coupling strength 
($\Gamma = 0.6$), the LFP temporal decorrelation becomes considerably slower 
and the envelope of the autocorrelogram displays amplitude modulations 
indicating that the LFP signal is quasi-periodic. In parallel, 
the gamma-band spectral plateau is replaced by a system of narrow peaks at 
incommensurate frequencies. 
The raster plot of activity (not shown) continues to display a temporally irregular 
oscillation; however spatial fluctuations in the width of consecutive bumps of spiking 
activity are reduced with respect to the fully-coupled case.
For further reduction of the interlayer coupling to $\Gamma = 0.2$, 
the LFP autocorrelogram starts revealing periodicity of the signal over long 
time scales. The multiple narrow spectral resonances merge into a single 
prominent resonance in the gamma-band and secondary harmonic peaks also appear.
Finally, for $\Gamma=0$ (Figure~\ref{fig:Decoupling}E, F), the LFPs are still 
substantially autocorrelated after several 
hundredths of ms. Spectra in the synchronous regime are harmonic at any contrast level. 
More details about the high contrast regime for completely uncoupled layers 
are presented in Appendix S7. 

Interestingly, qualitative modifications of the population dynamics when 
$\Gamma$ is varied do not occur in the low contrast regime, in which 
collective oscillations do not develop. As a matter of fact, independently of the coupling strength $\Gamma$, induced activity is asynchronous. 
Spiking and LFP responses to a low contrast stimulus between completely uncoupled or fully coupled layers are practically indistinguishable (not shown).

\subsection*{Stimulus repetition and chaotic sensitivity to initial conditions}

Up to now we have focused on the response of the network
to a time independent stimulus. Here we show that the inter-layer coupling
also strongly affects the response of the model hypercolumn induced 
by an external input which varies periodically, representing 
visual stimuli to V1 in the form of flashed or drifting gratings. 
In this situation, we characterize the neuronal responses by means of 
peristimulus time histograms (PSTHs) which express the probability of observing the 
firing of a spike at a given time relative to the onset of each stimulus 
presentation (see \textit{Methods} section). In the following, we focus on 
high contrast stimuli. 

The PSTH for $\Gamma=1$ is shown in Figure~\ref{fig:PSTH}A.
At the onset of the stimulus the probability of firing increases sharply, 
followed by a transient phase of reduced firing. {This feature is not evident in experimental PSTHs. It is due to the strongly synchronous recruitment of recurrent inhibition which follows the initial burst of activity, triggered by the rise of external inputs (instantaneous in our model). Notwithstanding, after a few tenths of a ms the firing probability rises again and remains then almost constant. }
This reflects the fact that the population responses are highly variable 
across trials as is clear in Figure~\ref{fig:PSTH}B. 
In each trial the response of the network consists of a sequence 
of episodes in which the neurons tend to fire together.
However, there are substantial trial-to-trial fluctuations
in the timing of these episodes and their amplitude 
(i.e. the numbers of recruited cells). Consequently,
although the presentations of the stimulus do give rise to 
synchronous activity, the PSTH histogram averaged over many trials is almost
flat after a peri-stimulus time on the order of the short
temporal decorrelation time of the induced oscillation. 

In contrast, for fully decoupled layers ($\Gamma = 0$), the 
PSTH averaged over many trials exhibits a long-lasting, 
although damped population oscillation, as plotted
in Figure~\ref{fig:PSTH}C. This is because when the layers are
decoupled the oscillations generated inside the layers
are close to being periodic and they maintain coherence over several
hundred milliseconds. Hence the timing  
of the oscillations does not fluctuate much across trials 
(Figure~\ref{fig:PSTH}D). Population oscillations are thus 
masked by averaging across multiple stimulus repetitions
only after many cycles.

The large trial-to-trial variability displayed by the network for $\Gamma=1$ (Figure~\ref{fig:PSTH}C) indicates a strong sensitivity to initial conditions (i.e. the network configuration at the onset of the stimulus). To further illustrate this sensitivity, we perturb the dynamics of the system by omitting artificially a single spike in a single neuron (out of $N=10000$) at the center of the bump of induced activity and we compare then the perturbed and the unperturbed dynamics. The results of this numerical simulation are illustrated by Figure~\ref{fig:Chaos}. As visible from the raster plot (Figure~\ref{fig:Chaos}A) and the population rate histogram (Figure~\ref{fig:Chaos}B) of the upper layer induced activity (at full contrast), the perturbed and the unperturbed collective oscillations can be distinguished  already after one oscillation cycle. After a few cycles, they have completely diverged. {Such} extreme sensitivity to perturbations or initial conditions is strongly indicative of dynamical chaos \cite{Schuster2005}. The sequence of states observed in our model for decreasing $\Gamma$ (from irregular to quasi-periodic to periodic, see Figure~\ref{fig:Decoupling}C,D) also suggests that chaos might emerge for strong inter-layer coupling and that its onset might occur according to a quasi-periodic scenario \cite{Newhouse1978, Schuster2005}. This is indeed one of the possible scenarios for the transition to chaos occurring in a related rate model \cite{Battaglia2007}. {As we discuss in detail in the Appendix S10, the chaotic nature of the dynamics of the network for $\Gamma=1$ and high contrast stimuli can be assessed by an estimation of its largest Lyapunov exponent $\lambda_{max}$ \cite{Schuster2005}. A positive value of this Lyapunov exponent is the manifestation of deterministic chaos, denoting exponentially fast separation of trajectories. Using techniques of non-linear time-series analysis \cite{Kantz2004} applied to very long stationary time-series of LFP from our model (see \textit{Methods} section and Appendix S10), we obtain the estimate $\lambda_{max} = 2.2 \pm 0.6$ ms$^{-1}$, which is indeed positive. Interestingly, the dynamics of the network with uncoupled layers ($\Gamma = 0$) fails to display a positive Lyapunov exponent (see Appendix S10), and it is therefore non chaotic, confirming the role of inter-layer coupling in inducing} (see also the \textit{Discussion} section). 

\section*{Discussion}

\subsection*{The structure of the model}

\subsubsection*{Multi-layer architecture}

The reduction of the full multi-layer structure of primary visual cortex 
(a cartoon of which is shown in Figure~\ref{fig:Circuit}A) 
to a simpler two-layer network (Figure~\ref{fig:Circuit}B) is 
a drastic simplification. 
Throughout this paper, we have emphasized that the 
two main cortical thalamo-recipient layers, i.e. IV and VI 
\cite{BlasdelLund1983, Ferster1996, Binzegger2004} are included within distinct model layers, 
corresponding respectively to the upper and the lower ring in our network architecture. We do not include separate rings for each of the six cortical layers.
However, in order to reflect the poly-synaptic nature of the
pathway from cortical layer IV to VI ---passing through
layers II/III and V \cite{Raizada2003, Thomson2003, Binzegger2004, Sincich2005,
Hirsch2006}--- we have made the latency of the connections from the upper
to the lower model layer larger than for the connections from the lower
to the upper model layer. The incorporation of additional layers within our 
model is in principle possible, but at the price of increasing further an 
already large number of parameters. Our choice of introducing just two 
layers  was guided by the need to keep the model as simple as possible, while retaining a multi-layer structure.

In the simulations described above, the external drive is smaller to the lower 
layer than to the upper layer. This choice was motivated by the fact that
thalamic projections toward  layer IV are more numerous than toward layer 
VI \cite{Binzegger2004}. Nevertheless, it should be noted that 
layer VI neurons have dendritic arborizations extending into layer IV where 
they can receive additional thalamo-cortical inputs \cite{Katz1987}.
However, as illustrated in Appendix S8, the behavior of the network remains qualitatively the same, 
if one adopts identical external drives for the two layers. {A second aspect that we have neglected about differences in the external drive to different layers, is the fact that the size of receptive fields depends on laminar location. In particular the receptive fields of layer VI neurons can be larger than the ones of layer IV neurons \cite{Gilbert1977, Martinez2005}. However, a proper description of the stimulus-size dependence of the inputs would require as well to take into account horizontal interactions between different layer IV  receptive fields fitting into a same larger layer VI receptive field, a modeling aspect that we hope to address in future investigations.}

\subsubsection*{Connectivity}

In our model intra-layer excitation is modulated more strongly with angular distance than intra-layer inhibition. However, the probability of 
inhibitory connections is larger than the probability of excitatory connection at any angular distance (Figure~\ref{fig:Circuit}E 
and Table~\ref{tab:ProbConn} for details). In addition, we choose conductance 
parameters such that individual inhibitory PSPs are stronger than excitatory
PSPs \cite{Holmgren2003}. Thus, intra-layer inhibition dominates intra-layer excitation at any distance. As a consequence, in the regimes 
explored in this paper, recurrent interactions are not sufficient to generate a tuned response by themselves.  However they sharpen the tuning already present in 
the spatially-patterned feed-forward LGN input. We use in the model probabilities of connection compatible with the wide ranges reported by \cite{Thomson2002, Yoshimura2005}. Other studies, like \cite{Holmgren2003}, find a larger probability of inhibitory connection. We verified however that the qualitative properties of the induced regimes of activity are preserved when inhibitory connections are consistently densified (see Appendix S8).

The dominantly inhibitory nature of mutual local interactions is essential
in our model for the emergence of prominent collective oscillatory behaviors 
in our network.  Oscillations are generated by mutual delayed interactions between inhibitory neurons, according to a standard mechanism already described in 
\cite{Brunel1999, Brunel2003, Brunel2006, Brunel2008, Wang2010}. In our model, excitatory neurons are not required for the generation of oscillations. Excitatory neurons are entrained by the oscillation paced by inhibitory cells. Indeed, if the activity of excitatory neurons is completely suppressed, or if synapses from excitatory to inhibitory neurons are removed, while increasing the drive to inhibitory neurons in order to maintain their rate of activity unchanged, the oscillations continue to exist and their frequency increases of less than five percent (see Appendix S8). We mention here that an alternative scenario exists in which the inhibitory-to-excitatory-to-inhibitory neurons feedback loop plays an active role in the generation of synchronous oscillations \cite{Brunel2000, Brunel 2003, Hansel2003, Geisler2005, Wang2010}. In this scenario delayed inhibitory feedback is still the cause of the oscillation, but the delay arise from the disynaptic nature of effective mutually inhibitory interactions, leading to a slower collective frequency. However, the analysis conducted in Appendix S8 clarifies that the scenario implemented in our model relies primarily on inhibitory interneurons alone.

Inter-layer connections in our model are as dense as intra-layer connections, 
but inter-layer excitation is more sharply modulated than intra-layer 
excitation. This results in a smooth arrangement of vertical excitatory 
synapses reminiscent of the organization of cortex into a continuum of 
anatomical columns without rigid boundaries \cite{Horton2005}.
This arrangement is critical for the fast temporal decorrelation 
of induced oscillations at high contrast (see below).  

Whereas the net inter-layer coupling is moderately excitatory in a 
local center, it is inhibitory in the  surround, as a combined effect of the 
broad profile of inter-layer inhibition and of the fact that lower-to-upper 
excitation toward inhibitory neurons (i.e. disynaptic inhibition) is less 
sharply modulated than lower-to-upper layer excitation toward excitatory 
neurons.  This is required in our model to account for the increase in
mean firing rate observed in layer inactivation experiments
\cite{Allison1994} (case $\Gamma = 0$ in our model).

\subsection*{The low and high contrast regimes}

Most of the simulations described above were performed in networks 
with a significantly smaller number of neurons ($N_E =4000$ excitatory neurons and $N_I=1000$ 
inhibitory neurons per layer)  than in a real hypercolumn in V1. However, we checked that our results are 
robust  against increases in network size. In particular,
this is the case for the existence of two dynamical regimes 
induced respectively by low  and high contrast stimulations
and for the two distinct mechanisms underlying the fast temporal 
decorrelation and broad-band spectral modulations in these two regimes. 

In the low contrast regime, the dynamics are asynchronous. 
However, the network tends to resonate at 
a specific frequency, producing an increase of power in the 
gamma frequency band, without developing stable oscillations. 
Weakly coherent oscillatory modes are excited only transiently by local
noise and then quickly damped.

On the other hand, in the high contrast regime the network 
activity is synchronous. However the collective rhythm undergoes  
random variations in the time interval between consecutive activity episodes 
in the network. This temporal irregularity 
is not due to local noise {(note that, in our model, recurrent inputs dominate over feed-forward inputs at low as well as at full contrast)}. It is produced intrinsically by the dynamics
by virtue of the interaction between distinct oscillating populations 
localized in the two subnetworks representing different 
depths in the cortical section. This results in rapid temporal decorrelation
of the induced activity.

The contrast at which the transition between these
two regimes takes place depends on the strength of fluctuations in the background noise.
For our choice of parameters, the transition occurs for $C \approx 10\%$. However, as discussed in detail in Appendix S9, if the variance in the LGN input current is increased 
consistently without changing its mean value, the transition can occur for an external drive, which is so large that it 
cannot be reached even for stimuli at full contrast. In such a condition, the induced activity is still asynchronous at 
high contrast and only transient oscillations can be detected, as in the 
recent modeling study by Mazzoni et al. \cite{Mazzoni2008}. 

It has been observed experimentally that the gamma-band synchronization of membrane potential fluctuations of nearby cells in V1 is larger in visually-induced activity than in spontaneous activity. Furthermore it is sustained over long stimulation durations, independently from stimulus properties or from the simultaneous observation of synchronized spiking activity. This leads to voltage crosscorrelograms with a manifest oscillatory component at gamma-range frequencies, damped quickly within only two or three oscillation cycles \cite{Yu2010}. These observations are compatible with the occurrence of a transition between an asynchronous low contrast regime and a synchronous high contrast regime. Indeed, pairwise CCos between membrane potentials are small in the low contrast regime (Figures~\ref{fig:LowContrast}D and~\ref{fig:PairwiseCCs}B), but large in the high contrast regime (Figure~\ref{fig:HighContrast}D and~\ref{fig:PairwiseCCs}D), even if spike CCos are always small, in agreement with many experimental reports \cite{Maldonado2000b, Zhou2008, Toyama1981, Schwarz1991, Hata1991, Ghose1994,  Das1995, Montani2007}.
{We remark that if the dynamics at high contrast would be asynchronous as the dynamics in absence of stimuli or for low contrast stimuli, then the pairwise crosscorrelations of \textit{both} spikes and voltages should be weak. Therefore, the coexistence of weak correlations between spikes with stronger correlations between membrane potentials (displaying furthermore a damping oscillatory component) is suggestive of the existence of a synchronous, rather than of an asynchronous, regime. The dynamics at high contrast of our model, characterized by irregular spiking (leading to weak spike crosscorrelations) and by temporally irregular collective oscillations (leading to quickly damped oscillatory voltage crossocorrelograms) is therefore compatible qualitatively with the experimental regime observed in \cite{Yu2010}. Conversely, this compatibility could not be claimed for the other two types of induced dynamics that our model can generate at full contrast, i.e. asynchronous, in the case of a large variance noise, or synchronous but approximately periodic (and therefore too slowly decorrelating), in the case of suppressed inter-layer interactions ($\Gamma = 0$).}

\subsection*{Synchronous chaos underlies the temporal decorrelation of 
the network collective oscillations in the high contrast regime} 

The rapid loss of temporal coherence of the synchronous induced activity
at high contrast is a remarkable property of our model.
Features of the model such as inter-layer inhibition, asymmetric 
interaction latencies in the lower-to-upper or in the upper-to-lower direction 
or different LGN driving levels to the different layers are not required for
this decoherence to occur. In contrast, the strong local inhibition
responsible for the local generation of the rhythm within each layer 
and  the net excitatory interactions between neurons in
close vertical alignment are crucial for this to occur.
In fact, if the inter-layer excitation profile is altered
by suppressing its modulation with orientation distance while keeping its
average strength constant, the decorrelation does not take place
(see Appendix S8). 

A similar mechanism  underlies the temporal decorrelation of synchronous 
oscillations in the  network models studied by \cite{Roxin2005, Roxin2006, Battaglia2007}.
These papers showed that collective oscillations induced in two populations
of neurons by local delayed inhibitory feedback can lose coherence when the 
two populations interact in an excitatory manner.
In \cite{Battaglia2007}, we studied a rate model consisting
of two networks, each composed of one excitatory and one inhibitory
populations.  Each of the networks was able to sustain synchronous
oscillatory activity by virtue of the local inhibition.
We computed the maximum Lyapunov exponent of the system
(see e.g. \cite{Schuster2005}) to show
that it undergoes a transition to a chaotic dynamical
state when the two networks are coupled by sufficiently strong excitatory
connections. In this state the network displays synchronous
activity, but instead of being periodic, the temporal variations
of the network activity are chaotic and thus the oscillations that
the network tends to develop lose temporal coherence within a few cycles.
A network operating in such a regime is said to be in a {\it synchronous
chaotic} state. In \cite{Roxin2005, Roxin2006} a single ring network with strong local inhibition 
was considered. The decoherence of the oscillations
occurred as the network underwent a spontaneous clustering into 
groups of oscillating neurons effectively interacting in an excitatory manner.  

{In agreement with the positivity of its largest Lyapunov exponent, also} the dynamics of our hypercolumn model in the high contrast regime {displays typical features of chaos}: exponentially fast damping of the local oscillations 
autocorrelograms (Figure~\ref{fig:LFPAC}C,D), spreading of the 
oscillation-related power over an extended continuous interval 
(Figure~\ref{fig:LFP}), 
and extreme sensitivity to initial conditions (Figure~\ref{fig:PSTH}B and Figure~\ref{fig:Chaos}).   
Therefore the decoherence of the population activity
which occurs at high contrast stems in the present model 
from the fact that the network operates in a synchronous 
chaotic regime. {We cannot exclude, obviously, that other mechanisms are contributing to the decorrelation of synchronous cortical oscillations. We stress nevertheless that such a global decorrelation, characterized by the coexistence of elevated instantaneous synchrony and fast loss of collective phase autocoherence, could not be induced by local external noisy inputs, unless they are spatially correlated over a range matching the size of the local circuit which generates the ongoing soscillation.}

We also conjecture that the underlying mechanism
of synchronous chaos is very general as it occurs in models in which
neurons are described in term of rate, integrate-and-fire
or conductance-based dynamics, with a simplified as well as more complex
multi-layer network architecture. We also conjecture that a similar mechanism
should act in even more realistic models, incorporating for instance a
two-dimensional spatial structure, similarly to the one used in \cite{Tao2004,
Tao2006}, provided that local inhibition is strong enough
to induce local oscillations and that excitation couples these local oscillators
at a longer range.

\subsection*{Comparison with previous works}

Chaotic dynamics as well as stable chaotic-like dynamics can occur in asynchronous states of activity \cite{Sompolinsky1988, Hansel1993, vanVreeswijk1996, vanVreeswijk1998, Marre2009, Zillmer2009, Jahnke2009, Sussillo2009, Rajan2010}. In this cases, the network dynamics explores a high-dimensional manifold in the phase-space, while, in our model, the irregular sparse firing of many neurons give rise to collective synchronous chaos (SC) with a lower dimensionality \cite{Shibata1998, Pecora1999, Kaneko2001} {(the fractal dimension of the chaotic attractor is likely to be smaller than five, as discussed in the Appendix S10).}

SC has also been found in previous models of local circuits in V1 
which consisted of only one single network with a ring architecture.
The model studied by Hansel and Sompolinsky in \cite{Hansel1992} considered 
one neuronal population coupled with excitatory instantaneous synapses.
It displayed a SC state in some appropriate range of parameters.
However, in this model, SC was  sensitive to the incorporation of synaptic 
time constants since it was destroyed with the introduction of synaptic time constants as small as 0.5 ms. 
The model by the same authors considered in \cite{Hansel1996} 
considered two populations of neurons, one excitatory and one inhibitory, 
coupled via synapses with realistic synaptic time constants. 
The dynamics of the neurons were based on a Hodgkin-Huxley type model 
with several cellular and synaptic conductances. The pattern of connectivity 
had a ``Mexican hat"  with local excitation and broad range inhibition. 
Numerical simulations of the model showed that in an appropriate parameter
range, the network settled in a SC state, characterized by strong temporal 
variability of the neural activity which was correlated across the hypercolumn.

In both of these models, the SC state was characterized by strong
neuronal pairwise spike correlations and wide variability in the 
firing of individual neurons which was induced by the chaotic nature of the 
population activity. This is essentially different from what happens in our 
two layers hypercolumn, in which, in the SC state at high contrast, the spike pairwise correlations 
are only slightly larger than in the low contrast asynchronous state, whereas 
the degree of irregularity in the spike trains are similarly large in both 
states ($CV \approx 0.9$). As a matter of fact, in the present model, the spike train irregularities
are mostly due to the local noise generated by the external inputs and to a 
lesser extent by the internal dynamics. Voltage CCos are large due to macroscopically correlated chaotic sub-threshold fluctuations, but spike CCos are still small.
Another essential difference is that 
in \cite{Hansel1996} the excitation was local and inhibition was broad, whereas 
the opposite is required in the present model, as well as in the single ring 
model in \cite{Roxin2006}. Last but not least, it is not clear to what extent 
the chaotic dynamics found in \cite{Hansel1996, Hansel1992} were specific to the 
model adopted there for the single neuron dynamics.

\subsection*{Predictions and perspectives}

The increase in synchrony of the activity with the contrast displayed by our model  
is in agreement with experimental results reported recently
in monkey V1 \cite{Henrie2005, Ray2010}.
More generally we should expect that varying a feature of a stimulus in
a way that increases the external drive on V1 network should have 
a similar effect. This is consistent with other recent results showing that 
varying the size of a visual stimulus \cite{Gieselmann2008} or attention \cite{Fries2001, Fries2008} 
strengthens the coherence in the activity of V1 neurons. 

In the low and large contrast regimes identified in our model the increased 
gamma power in the LFP spectra is broadband. 
At low contrast, the loss of coherence of the oscillations 
in the LFP in a few tenths of a milliseconds is due to noise. At large contrast, 
it is a consequence of the chaoticity of the LFP time-series. The behavior 
of our model in both these regimes is compatible with recent results by 
Burns et al \cite{Burns2010}, because of its lack of sustained auto-coherence 
of induced oscillations.  

Our simulations predict that infra-granular layer inactivation should 
globally affect the experimentally observed spectral properties of induced LFPs by 
enhancing its periodicity. Single-layer inactivation experiments based on 
pharmacological or local cooling techniques \cite{Bolz1986, Schwark1986, 
Allison1994} or with optogenetic techniques  \cite{Cardin2009, Sohal2009} 
might be used to test this prediction. {Furthermore, manipulations in which the firing of a single additional spike is induced (or suppressed, analogously to the simulation of Figure~\ref{fig:Chaos}) can be performed. Extreme sensitivity to single-spike perturbations was experimentally proved in the case of asynchronous spontaneous cortical dynamics \cite{London2010}. It would be interesting to repeat similar experiments in a stimulus-induced regime of oscillatory activity, in order to study the impact of the addition of a single spike on the time-course of ongoing LFPs.}

In the present study we focused on the role of the interactions
between cortical layers in promoting temporal decoherence
of gamma oscillations via the generation of synchronous chaos 
in a network with the size of a typical classical receptive field in V1. 
It would be interesting to investigate whether horizontal interactions 
which extend at distances beyond the  
classical receptive field also contribute to the loss of temporal
coherence via a similar mechanism when the visual stimuli
are extended. The basic two-ring network developed in this paper 
can be replicated into a bi-dimensional architecture including 
long-range excitatory interactions in order to investigate this potential
contribution. This framework can be also applied to assess how the phase
relationship between activity at different locations 
in V1 (e.g. between center and surround of an extended stimulus)
depend on the polarity of long range interactions. {Furthermore, an additional source of decorrelation might be inter-areal interactions occurring at an even longer range.}

Finally, we have here considered temporal decorrelation induced by excitatory interactions between populations oscillating due to delayed mutual inhibition. It would be interesting to investigate whether a similar decorrelation phenomenon can arise when the mechanism for the local generation of oscillations is different, and is based for instance on circuit loops with active involvement of pyramidal cells \cite{Wilson1972, Li1989, Brunel2000, Brunel 2003, Hansel2003, Geisler2005, Wang2010}.

\section*{Methods}

Our model of a functional hypercolumn in V1
consists of two interacting rings of neurons, an upper and a lower ring,
each comprising $N_E$ excitatory and $N_I$ inhibitory neurons
connected recurrently. We denote by $N= 2 (N_E+N_I)$ the total number of 
neurons in the network.  Each neuron is labeled by its location on the ring 
to which it belongs; i.e. by an angular coordinate $\vartheta$, ranging 
conventionally from -90 to +90 degrees \cite{BenYishai1995, Hansel1998}.
All the neurons receive an external input composed of 
two contributions. One represents the LGN input to V1. It depends on two 
parameters $C$ and $\theta_{stim}$ corresponding to the contrast and the 
orientation of a visual stimulus. The other contribution accounts for the
background inputs $V1$ receives from subcortical regions.

\subsection*{Single neuron dynamics}

Throughout the paper, we use single-compartment Exponential Integrate-and-Fire 
model neurons (EIF;  \cite{Fourcaud2003}). In this model
the membrane potential $V$ is given by the equation:
\begin{equation}\label{EquationEIF}
\frac{dV}{dt} = -\frac{1}{\tau_m}(V-V_L)+\psi(V)+\frac{I_{syn}(t)}{C}
\end{equation}
where $C$ is the membrane capacitance,  $\tau_m$ the membrane time-constant, 
$V_L$ the leak potential,  $I_{syn}$  the total synaptic input current 
to the neuron. The function $\psi(V)$ is: 
\begin{equation}\label{EquationPsiEIF}
\psi(V) = \frac{\Delta_T}{\tau_m}\exp\left(\frac{V-V_T}{\Delta_T}\right)
\end{equation}
For a constant input above a threshold current ($\sim 0.113$ nA for the parameters adopted here)
the solution of \eqref{EquationEIF} diverges to infinity in finite time.
This divergence is identified with the firing of a spike. 
The parameters $\Delta_T$ and $V_T$ characterize how sharp the 
initiation of the spike is and the voltage at which it occurs. The spike downswing is not explicitly modeled. 
After each spike event, the voltage needs to be reset. 
A refractory period must then follow. 

We model this refractoriness in a different way for excitatory and inhibitory neurons.
In the case of excitatory neurons, following the emission of a spike at 
time $t_{spike}$, the parameters
$\tau_m$, $V_T$ and $V_L$ are updated according to the equations
\cite{Badel2008a} , 
\begin{equation}\label{EquationDynamicLeak}
V_L = V_L^0 + A_{V_L}\exp\left(-\frac{t-t_{spike}}{\tau_{A, V_L}}\right) -   B_{V_L}\exp\left(-\frac{t-t_{spike}}{\tau_{B, V_L}}\right)
\end{equation}
\begin{equation}\label{EquationDynamicThreshold}
V_T = V_T^0 + A_{V_T}\exp\left(-\frac{t-t_{spike}}{\tau_{A, V_T}}\right)
\end{equation}
\begin{equation}\label{EquationDynamicTau}
\frac{1}{\tau_m} = \frac{1}{\tau_m^0} + A_{\tau_m}\exp\left(-\frac{t-t_{spike}}{\tau_{A, \tau_m}}\right)
\end{equation}
The membrane potential is reset to a value $V_{reset}$ which is sub-threshold. Furthermore $V_T$ is strongly depolarized after a spike. Therefore the event that two spikes are closely emitted in time by a same neuron is extremely unlikely and, in practice, never occurs.

For inhibitory interneurons, we use a ``hard'' refractory period instead, 
suspending the numerical integration for a time $\tau_{ref}$ after voltage 
reset \cite{Fourcaud2003}. Therefore,  $V_L=V_L^0$, $V_T=V_T^0$ and 
$\tau_m=\tau_m^0$.

Parameters for excitatory neurons are chosen to coincide with fits of 
pyramidal neurons traces, following \cite{Badel2008a}. We use analogous 
parameters for inhibitory neurons, apart from halved membrane capacitance 
and time constant $\tau_m$, consistent with experimental evidence 
\cite{McCormick1985} and fits of interneuronal traces presented in 
\cite{Badel2008b}. All single neuron parameters are given in 
Tables~\ref{tab:SingleNeuronStatic} and~\ref{tab:SingleNeuronDynamic}. 

\subsection*{The synapses}

We use three kinds of synaptic currents, modeling 
inhibitory (GABA-type), fast excitatory (AMPA-type) and slow excitatory 
(NMDA-type) synaptic inputs. No voltage dependence is introduced for the 
parameters of the slow excitatory synaptic current. A spike in an inhibitory 
pre-synaptic neuron evokes a GABA-type post-synaptic potential (PSP) in all 
the post-synaptic neurons; a spike in an excitatory presynaptic neurons evokes 
composite AMPA- and NMDA-type PSPs. 

The synaptic current produced by a single incoming spike is described 
as $I_{syn, spike}(t) = -g_{syn}(V-V_{syn})s(t)$,  
where $g_{syn}$ is the 
peak synaptic conductance, $V_{syn}$ the reversal potential of the 
synapse ($V_{AMPA} = V_{NMDA} = 0.0$ mV, $V_{GABA}  = -75$ mV). Denoting 
as $t_{spike}$ the time of pre-synaptic firing and with $d$ the synaptic 
latency, the function $s(t)$ is: 
\begin{equation}
s(t) = \frac{1}{\mathcal{N}}\left[e^{-\frac{t-(t_{spike}+d)}{\tau_{d}}} - e^{-\frac{t-(t_{spike}+d)}{\tau_{r}}}\right]
\end{equation}
where the constant $\mathcal{N}$ is such that it normalizes to unity the peak 
of $s(t)$. All the synaptic conductances in the network are calibrated 
to give unitary PSPs at resting potential in a range compatible with 
experimental observations \cite{Holmgren2003}. 

The values of the synaptic times and synaptic peak conductances are 
given in Table~\ref{tab:Condu}, for a network including $N_E = 4000$ 
excitatory neurons and $N_I = 1000$ inhibitory neurons per layer. 
Synaptic peak conductances are rescaled for larger 
networks, according to Eqs. \eqref{eq:rescale} and \eqref{eq:rescale2}. 
Synaptic latencies are given in Table~\ref{tab:Latencies}. 

\subsection*{Network connectivity} 

Each of the two layers of the hypercolumn is modeled by a ring-network 
\cite{BenYishai1995, Somers1995, Hansel1996, Hansel1998}. 
Unless specified otherwise, the simulations described in 
this paper were performed 
for a network comprising $N_E = 4000$ excitatory cells 
and $N_I = 1000$ inhibitory cells per ring, for a total 
of $N = 2(N_E + N_I) = 10000$ neurons in the hypercolumn. 
{Note that a very similar network architecture was used in 
\cite{Ardid2007, Ardid2010} but with a completely different interpretation.}

Intra-layer and inter-layer excitatory and inhibitory connections are 
random. The  probability of connection between two neurons is spatially 
modulated and depends on the angular coordinates $\vartheta_{pre}$ 
and $\vartheta_{post}$ of the pre- and post-synaptic neurons. It also 
depends on the nature (excitatory or inhibitory) of pre- and post-synaptic 
cells and on their absolute (lower or upper layer) and relative 
(intra-layer or inter-layer) depth. All the profiles of connection 
probability are parameterized as:
\begin{equation}
P(\vartheta_{pre}, \vartheta_{post}) = \left[ p^{(0)} + 
p^{(1)}\cos 2(\vartheta_{pre} - \vartheta_{post})\right]_+
\end{equation}
Here, $[\cdot]_+$ denotes rectification; {i.e.} 
$[x]_+ = 0$ if $x < 0$, else $x]_+ = x$.
The probabilities of connection for intra-layer excitatory and inhibitory 
connection are identical for each of the two layers.

In order to study the scaling properties of the dynamics it is important 
to guarantee that the spatial mean and spatial fluctuations
of the time averaged recurrent synaptic 
currents received by each neuron are preserved when considering 
networks of different sizes. This requires 
a suitable modification of the probabilities of connection and of the peak 
synaptic conductances when passing from a network of size $N$ to a network 
of size $N'$ \cite{Golomb2000}. For an arbitrary peak recurrent synaptic 
conductance $g_x$, the probabilities of connection (and, correlatively the average number of pre-synaptic cells of each type) are scaled as:
\begin{equation}\label{eq:rescale}
\frac{1}{N'}\left(\frac{1}{P'_x}-1\right) = \frac{1}{N}\left(\frac{1}{P_x}-1\right)
\end{equation}
and peak conductances as:
\begin{equation}\label{eq:rescale2}
P_x N g_x =P'_x N' g'_x
\end{equation}
Here the index $x$ stands for different kinds of synaptic connections, each one potentially characterized by different mean probabilities of connections and connection strengths (i.e. originating from upper or lower layer excitatory or inhibitory neurons and directed toward upper or lower layer excitatory and inhibitory neurons).

Sizes between $N_E = 1000$ ---for a total network size of 
$N= 3000$ neurons--- and $N_E = 32000$  ---for a total 
network size of $N= 80000$ neurons--- are compared in 
scaling analysis of synchrony properties.
The  parameters for  $N_E = 4000$ and $N_I = 1000$ are 
given in Table~\ref{tab:ProbConn}. 
Probabilities of connection are compatible with the ranges reported by \cite{Thomson2002, Yoshimura2005}.

\subsection*{Model of the LGN input}

We assume that the firing rate of a single LGN neuron, $r(C)$ is 
related to the stimulus contrast, $C$, ($C = 1 \div 100$ \% )
by the equation  \cite{Somers1995}:  
\begin{equation}\label{contraequa}
r(C)=  r_0+ r_1\log_{10}(1+C) \mbox{\,\, Hz}
\end{equation}
where $r_0$ is the spontaneous activity of the neuron in dark conditions.
Subsequently, we model the LGN input to a cortical cell as an 
AMPA-type synaptic connection with peak conductance $g_{LGN}$, driven by
homogeneous Poisson spike trains
with rate  $R_{LGN}(\vartheta, \vartheta_{stim}, C)$,
\begin{equation}\label{eq:TunLGN_1}
R_{LGN}(\vartheta, \vartheta_{stim}, C) = R_0+ \left [ R_1(C)\left(1-\epsilon + \epsilon\cos 2(\vartheta - \vartheta_{stim})\right)\right]_+
\end{equation}
with:
\begin{equation}\label{eq:TunLGN_2}
R_1(C)=\bar{R}_1 \log_{10}(1+C)
\end{equation}
Here the parameter $\epsilon$ controls the broadness of tuning of the
LGN input. It is  set to 1 in all our simulations. 
Note that $R_{LGN}$ is maximum when  $\vartheta_{stim} = \vartheta$.
The contrast $C$ and, correspondingly, the term $R_1(C)$ can also be time-dependent (see later section on peristimulus time histograms).
The LGN input targets both layers. There is anatomical evidence that
thalamo-cortical synapses target mainly layer IV and to a lesser extent
layer VI \cite{BlasdelLund1983, Binzegger2004}. Accordingly, in all the 
simulations presented in this paper, $g_{LGN}$ in the lower layer is smaller 
by a factor of two than in the upper layer. Parameters describing LGN input 
properties are given in Table~\ref{tab:Contrast}. 

{For the adopted parameters, feed-forward inputs from LGN never dominate over recurrent inputs from the two layers of the network, consistently with the massively larger number of cortico-cortical synapses than thalamo-cortical synapses in the primary visual cortex \cite{Binzegger2004}.
The relative weight of feed-forward inputs with respect to recurrent inputs depends on contrast, doubling in our model from no more than 20\% at low contrast to no more than 40\% at high contrast stimulation (not shown). 
}

An alternative parameter choice for the tuned component of the LGN input, leading to noisy input current with a larger variance, is analyzed in Appendix S9.
For the noisy inputs used in this paper as well as for the ones used in Appendix S9, the resulting sub-threshold voltage fluctuations are on the order of $\sim$6 mV at full contrast, compatible with experimentally observed fluctuation ranges  \cite{Anderson2000b, Cardin2007}. Voltage fluctuations are comparable in the two regimes, because the increase in amplitude of external input current fluctuations is paralleled by a decrease in amplitude of net input conductance fluctuations, due to reduced synchrony among the recurrent inputs (see Appendix S9).

More details about the mapping from stimulus contrast to input rates can be found in Appendix \textbf{S11}.

\subsection*{Background cortical noise}

In addition to the LGN input, excitatory and inhibitory cells are driven by 
an untuned noisy input, representing the background firing of other cortical 
areas. This input is  modeled by a single AMPA-type synapse per cell, with 
peak conductance $g_{bg}$ activated by Ornstein-Uhlenbeck processes 
\cite{Rudolph2004}. Input spikes are generated independently for each cell;
however all the cells share the same instantaneous 
input rate $R_{bg}(t)$ obeying the stochastic differential equation:
\begin{equation}\label{eq:OU}
\frac{d R_{bg}(t)}{dt} = -\frac{1}{\tau_{bg}}(R_{bg}(t) - \mu_{bg}) + \sqrt{\frac{2\sigma_{bg}^2}{\tau_{bg}}}\,\xi(t)
\end{equation}
where $\xi(t)$ stands for Gaussian white noise and $\mu_{bg}$ is the mean, $\sigma_{bg}$ the volatility and $\tau_{bg}$
the filtering time-constant of the stochastic process. Parameters are given in 
Table~\ref{tab:Background}.

\subsection*{Numerical integration scheme}

The dynamical equations are integrated using a fourth-order
non-adaptive step Runge-Kutta scheme. Integration step was 0.2 ms. Because of the exponentially fast divergence
of the membrane in relation with firing, particular care is needed
to ensure the stability of the numerical integration of equation
(\ref{EquationEIF}).
Following  \cite{Fourcaud2003}, the numerical integration of the membrane
potential $V$ of a given neuron is stopped as soon as $V$ reaches a finite
cutoff voltage $V^{\star}$.
In our simulations, we use $V^{\star} = -30$ mV.
This choice ensures that the non-linear term $\psi(V^{\star})$ is the dominant
contribution to the neuronal currents for $V > V_{th}$. Under this condition,
the leakage and the synaptic currents can be neglected, making it possible to
compute analytically the time left before the actual divergence of the
potential. Assuming that the integration is stopped at $t=t_{stop}$
when $V = V_{stop} > V^{\star}$, the time of the next spike is given by
$t_{spike} \simeq t_{stop}+\tau_m e^{(V_T - V_{stop})/\Delta_T}$.
In addition, for our choice of $V^{\star}$, $t_{spike}-t_{stop}$ is large
compared to the integration-step $\Delta t$, thus avoiding numerical errors in
spike-time estimation due to the exponentially fast growth of $V$ in
proximity of the divergence. The membrane potential is then reset
to a value $V_{reset}$ immediately after a spike.

The Ornstein-Uhlenbeck process giving $R_{bg}(t)$ is computed using 
the properties of the exact solution to equation \eqref{eq:OU}. 
This means that $R_{bg}(t +\Delta t)$ is normally distributed with mean 
$\hat{m} = R_{bg}(t)e^{-\Delta t/\tau_{bg}} + 
\mu_{bg}(1-e^{-\Delta t/\tau_{bg}})$ and standard deviation 
$\hat{s} = \sigma_{bg}\sqrt{1-e^{-2\Delta t/\tau_{bg}}}$ \cite{Gillespie1996}.

\subsection*{Response tuning and contrast response function}

In order to study the tuning properties of the neuronal responses
we present stimuli at 12 different orientations 
$\vartheta_{stim}$ in an interval between -90 degrees and +90 degrees, 
at at least five different contrast values $C$ per each orientation. 

Tuning curves are derived for each neuron by measuring their average 
firing rate for each of the tested orientations and  
contrasts and are characterized by computing their skewness and their 
circular variance \cite{Ringach2002}. See Appendix S3 
for more details. Population average tuning curves are computed after rotating 
single neuron tuning curves so that their maximum is at $\vartheta=0$
(see Figure~\ref{fig:TuningCRF}). 

The contrast-response functions, CRF$(C)$, are computed for 
each neuron by measuring its peak firing rate ({i.e.} 
its firing response to a preferred orientation stimulus) at each given 
level of contrasts. Each individual 
CRF  is fitted to a hyperbolic ratio function \cite{Contreras2003}:
\begin{equation}
\mbox{CRF}(C) = R_{max}\frac{C^n}{C^n + C_{50}^n}
\end{equation}

\subsection*{Measures of synchrony}

To measure the degree of macroscopic synchrony in the steady state
of a network comprising an arbitrary number $N$ of neurons, we follow the 
method used in \cite{Hansel1996, Golomb2000}.
It is grounded on analysis of the temporal fluctuations of
macroscopic observables of the networks such as the
instantaneous activity or the instantaneous membrane potential
averaged over a population of neurons of size $K$.
For instance, for the latter, one evaluates at a given time, $t$, 
the quantity:
\begin{equation}
V(t) = \frac{1}{K} \sum_{i=1}^{K} V_i(t)
\label{s1}
\end{equation}
The variance of the time fluctuations of $V(t)$ is
\begin{equation}
\sigma_V = \left< \left[ V(t) \right]^2 \right> -
\left[ \left< V(t) \right> \right]^2 \label{s2}
\end{equation}
where $\left< \ldots \right> = \int_0^{T} dt \, \ldots$ denotes
time-averaging over a large time, $T$.
After normalization of  $\sigma_V$  to the average over the population of
the single cell membrane potentials:
\begin{equation}
\sigma_{V_i} = \left<\left[ V_i(t) \right]^2 \right> -
\left[ \left< V_i(t) \right> \right]^2 \label{average}
\end{equation}
one defines a {\it synchrony measure}, $\chi (K)$ by:
\begin{equation}
\chi^2 \left( K \right) = \frac{\sigma_V^2}{ \frac{1}{K} \sum_{i=1}^N
\sigma_{V_i}^2} \label{s3}
\end{equation}
This measure takes values between 0 and 1.
In the limit  $K \rightarrow \infty$  it behaves as:
\begin{equation}
\chi \left( K \right) = \chi \left( \infty \right) +
\frac{a}{\sqrt{K}} + O(\frac{1}{K}) \label{sig}
\end{equation}
where $a$ is some constant, between $0$ and
$1$.
In particular, $\chi (K) = 1$, if the system is fully synchronized (i.e.,
$V_i(t)=V(t)$ for all $i$), and $\chi \left( K \right) = O(1/\sqrt(K))$
if the state of the system is asynchronous.
Asynchronous and synchronous states
are unambiguously characterized in the thermodynamic limit (i.e., when
the number of neurons is infinite). In the asynchronous
state,  $\chi(\infty)=0$. By contrast, in synchronous states,
$\chi(\infty) >0$.

To characterize the degree of synchrony in the membrane potentials of 
neurons $i$ and $j$, we compute the cross-correlation function:
\begin{equation}\label{eq:Corr}
\mbox{CC}(V_i, V_j)[\tau] = \frac{ \big\langle(V_i(t)-\langle V_i(t)\rangle)\cdot (V_j(t+\tau)-
\langle V_j(t)\rangle) \big\rangle}{\sqrt{\sigma^2_{V_i(t)}\sigma^2_{V_j(t)}    }   }
\end{equation}
The value of the normalized cross-correlogram for zero time-lag gives the pairwise crosscorrelation coefficient (CCo):
\begin{equation}\label{eq:Corrbis}
\mbox{CCo}(V_i, V_j) = \mbox{CC}(V_i, V_j)[\tau=0]
\end{equation}
To estimate the degree of synchrony in the spiking activity of these neurons,
discrete spike trains are first convolved with a 
square window of width $B$,  thus generating a continuous spike-count signal. 
Equations~\eqref{eq:Corr} and~\eqref{eq:Corr}, with $V_i$ replaced by such smoothed spike trains,
is used to compute crosscorrelograms and CCos for spiking activities \cite{Renart2010}.
We use a smoothing window size of $B = 20$ ms.

CCos and crosscorrelograms are estimated over simulated recordings lasting 100 s of real time.
For CCos between membrane potentials only pairs of neurons within 
a 18$^\circ$ region centered on an angular coordinate matching the orientation 
of the presented stimulus are considered. In the case of spike trains, 
neurons in this region whose spike train contained fewer than 100 spikes are 
further excluded. Various stimulus orientations are pooled together to 
improve the estimation of CCo distributions. 

\subsection*{Local field potentials}

LFPs are believed to be an aggregate measure of the synaptic activity of 
several hundreds of neurons in the vicinity of the recording electrode 
\cite{Mitzdorf1985, Katzner2009}. To evaluate the LFP in a 
given site, we thus arbitrarily average input synaptic currents in a small 
angular sector of $9^\circ$ centered 
on the considered angular position. 
LFPs are estimated over neurons of the upper layer only, reflecting the 
fact that superficial neurons should supply the largest contribution to the 
signal recorded by an applied recording tip. For the normally used size 
of $N_E = 4000$ excitatory neurons and $N_I = 1000$ inhibitory neurons 
per layer, this corresponds to averaging over 200 excitatory and 50 
inhibitory upper layer neurons for each considered LFP recording site.

Autocorrelograms of the LFPs are computed as:
\begin{equation}
\mbox {AC}[\mbox{LFP}](\tau) = \langle \mbox{LFP}(t)\cdot \mbox{LFP}(t+\tau)\rangle -
\langle\mbox{LFP}(t)\rangle^2
\end{equation}
We evaluate non-normalized autocorrelograms, in such a way that the zero-lag value AC$[\mbox{LFP}](\tau = 0)$ measures the variance of the temporal
fluctuations of the LFP and has known size-scaling properties, which are different in synchronous and asynchronous regimes \cite{Golomb2000}.

Power spectra are computed using conventional FFT techniques, as the square 
modulus of the Fourier Transform of signal autocorrelation. 
Windowing is applied to LFP-like signal time-series to reduce unwanted frequency 
leakage, following the Welch method \cite{Oppenheim1999}). 
An additional moving average smoothing is applied for visualization 
purposes. We measure power in arbitrary 
logarithmic units. Since we are interested in qualitative analysis of the overall 
shape of the spectra rather than in absolute power estimations, for each considered regime we assign 
a unit value at the power at 0 Hz frequency for 0\% of contrast.

Autocorrelation and spectral analysis of LFP-like signals are based on 
time-series lasting 100 s of real time, with a sampling rate of 5 kHz.

\subsection*{Multi-unit-activity}

{
The MUA signal reflects the spiking activity of few neurons in the immediate vicinity of the recording electrode \cite{Buzsaki2004}. Typically, the recorded MUA is separated in only a small number of contributing single units \cite{Quiroga2004}. To evaluate MUA at a given site, following \cite{Ardid2010}, we sum together the spike trains of three randomly selected cells within a small angular sector of $9^\circ$ centered on the considered angular position (the same used for the evalutation of the LFP). This discrete signal is then transformed into a continuous signal by convolving it with a gaussian window (1 ms of variance).}

{
We compute then the coherence \cite{Mitra1999} between the LFP and the MUA at a same site by taking the modulus of the normalized product of their complex Fourier Transform, using the Welch method \cite{Oppenheim1999}, as in the case of the LFP power spectrum estimation:
\begin{equation}
C(f) = \frac{S_{LFP}(f)\cdot S_{MUA}^*(f)}{\sqrt{|S_{LFP}(f)|^2\cdot|S_{MUA}(f)|^2}}
\end{equation}
where $S_{LFP}(f)$ and $S_{MUA}(f)$ denote the Fourier transform of the autocorrelograms of the LFP and of the MUA signals, respectively, a star denotes complex conjugation and $|\cdot|$ complex absolute value. Such MUA-LFP coherence is a real quantity in the unit interval $0 \le C(f) \le 1$, and provides an absolute (linear) measure of the phase synchronization between the two signals in different frequency bands.
We average then the result over twenty different randomly chosen triplets of cells, in agreement with the experimental habit to average together different MUA recordings with only approximately similar selectivity properties \cite{Ardid2010}.

Note that both our MUA and our LFP signals are not designed to reproduce faithfully realistic electrophysiological signals. Due to the extreme sensitivity of the coherence measure \cite{Mitra1999}, therefore, we cannot expect the MUA-LFP coherence measure we compute to have nothing more than an illustrative purpose. 

}

\subsection*{Inter-layer coupling strength and layer decoupling}

Layer decoupling experiments are performed by multiplying the peak 
conductances of all the AMPA-type, NMDA-type and GABA-type synapses from 
the lower layer to the upper layer by a factor $\Gamma$ varying between 1 
and 0. A value of 1 corresponds to the case of fully-interacting layers, and 
a value of 0 corresponds to fully uncoupled layers.

For each excitatory neuron in the upper layer, at high contrast, the peak 
response after layer decoupling is compared with the peak response of the 
same neuron in the fully coupled network case. In comparing peak responses, 
we incorporate the fact that the tuning curves of many neurons change their 
preferred orientation or their skewness after full or partial lower layer 
inactivation.

\subsection*{Peristimulus time histograms}

To simulate the flashing of a grating, for a given network 
realization we perform numerical simulations in which the tuned LGN input rate is 
not constant. More specifically, this tuned component is still modeled according to equations \eqref{eq:TunLGN_1} and \eqref{eq:TunLGN_2}, but the contrast is now modulated in time:
\begin{equation}
C=C(t)=\left\{\begin{array}{ll}
0\% 		& 	n\cdot 1.5 \mbox{ s} \le t <   (n\cdot 1.5  + 0.5) \mbox{ s} \\
95\% 	&	(n\cdot 1.5 + 0.5) \mbox{ s} \le t < (n+1)\cdot 1.5 \mbox{ s}
\end{array}\right.\,\,\,\,
,\,\,\, n=0,1,2,\ldots
\end{equation}
Phases lasting 0.5 s in which $R_1 = 0.0$ are therefore alternated 
with phases lasting 1 s in which $R_1 = R_1(95\%)$, leading to a square wave 
time-course of the input LGN rate. We consider only cells whose preferred orientation falls within a sector $9^\circ$ wide centered on the orientation of the presented stimulus and we use four different stimulus orientations. For each of the 
orientations, the stimulus is flashed 1000 times. An overall sample of 800 cells (200 per orientation) is thus considered.
We count across stimulus repetitions and cells how many times a spike is emitted within 2 ms from a specified time following the onset of a stimulus.
The probability that a neuron will fire at a given time after stimulus presentation is then evaluated as (number of spikes in a time-from-stimulus bin) / (number of stimuli 
repetitions) / (number of sampled cells).

\subsection*{Single spike perturbation}

To study the sensitivity of induced dynamics to a small perturbation, we perform a simulation in which just a single spiking even is omitted and we compare it with the unperturbed simulation. We select a putative spiking time of a neuron whose preferred orientation matches the one of the applied 95\% of contrast stimulus. No spike is then sent to the post-synaptic targets, we only reset the potential and the other time-dependent parameters of the failing presynaptic neuron to their just-after-spike values. Precisely the same realizations for all the stochastic noisy input processes are taken for the unperturbed and the perturbed dynamics.

\subsection*{Estimation of the largest Lyapunov exponent}

{Rather than estimating the largest Lyapunov exponent through a ``direct'' method based on the application to the system of vanishingly small perturbations or through a ``exact'' method based on the integration of the linearized system \cite{Benettin1980}, we measure the largest Lyapunov exponent of the induced dynamics of the system at high contrast through a non-linear analysis of a long time-series (600 minutes of real time) of the associated LFP signal. For a thorough introduction and a rationale to the used methodologies the reader is invited to refer to textbooks like \cite{Kantz2004}. We try just here to give a flavor of the employed techniques. The first step is the construction of proper ``embeddings'' of this time-series. Given a discretely sampled time-series $\ell_t = \mbox{LFP}(t)$, a reconstruction delay $\tau$ and an embedding dimension $m$, we construct a new $m$-dimensional sequence:
\begin{equation}
\vec{\ell}_t = \left(\ell_{t-(m-1)\tau}, \ell_{t-(m-2)\tau}, \ldots, \ell_t\right)
\end{equation}
It can be proven \cite{Takens1981, Sauer1991} that the latter time-delay embedding provide in general a one-to-one image of the original phase-space attractor of the dynamics generating the measured time-series, provided that the used embedding dimension $m$ is large enough. The general idea of the method is then to identify by systematic search pairs of points $\vec{\ell}_t$ and $\vec{\ell}_{t'}$ which lie at a (euclidean) distance in the delay-embedding space smaller than a specified very small $\epsilon$. Such points are said to be \textit{neighbors}. It is therefore possible to consider the distance $\delta_0 = \|\vec{\ell}_t - \vec{\ell}_{t'}\|$ as a ``small perturbation'', which should grow exponentially in time if the dynamics is chaotic. The eventual divergence of the trajectories originating by neighbor points can be monitored by the quantity $\delta_k = \|\vec{\ell}_{t+k} - \vec{\ell}_{t'+k}\|$. If a time range for which $\delta_k \propto \delta_0\exp(\lambda k)$ than $\lambda$ cohincides with the maximum Lyapunov exponent $\lambda_{max}$ \cite{Rosenstein1993, Kantz1994}. 
}

{
We select a reconstruction delay of $\tau = 400$ ms, much larger than the decorrelation time of the induced LFP oscillation. The minimum embedding dimension for a consistent estimation of the largest Lyapunov exponent can be estimated by monitoring the fraction of ``false neighbors'' pairs, i.e. pairs of points that are neighbors in a $D$-dimensional embedding (due to a projection of the attractor to a space with a too small dimensionality) but that there are no more such in an embedding with a larger dimension $D'$ \cite{Kennel1992}. Such analysis, summarized in the Appendix S10, indicates a critical embedding dimension lower than five and probably larger than three (even if a precise estimation is difficult due to the presence of noise). Practically, we estimate the largest Lyapunov exponent by evaluating the quantity:
\begin{equation}
\ln\frac{\delta_k}{\delta_0}(\epsilon, m, k) = \left\langle \ln \left(\frac{1}{|\mathcal{U}_t(\epsilon)|}\sum_{\vec{\ell}_{t'}\,\in \,\mathcal{U}_t(\epsilon)} \|\vec{\ell}_{t+k} - \vec{\ell}_{t'+k}\|\right)\right\rangle_t
\end{equation}
for various $\epsilon$, $m$ and $k$, where $\mathcal{U}_t(\epsilon)$ is the set of points at a distance $d \le \epsilon$ from $\vec{\ell}_{t}$ and $\langle\cdot\rangle_t$ denotes average over time. If $\ln\frac{\delta_k}{\delta_0}(\epsilon, m, k)$ displays a linear increase in a reasonable range of $k$ with matching slopes for different sufficiently large $m$ and for few decades of $\epsilon$, than the average slope of the linear sections of $\ln\frac{\delta_k}{\delta_0}(\epsilon, m, k)$ provides a robust estimation of $\lambda_{max}$. More details on our estimation of $\lambda_{max}$ for the high contrast induced LFPs for $\Gamma = 1$ and $\Gamma = 0$ are given in the Appendix S10.
}

\section*{Acknowledgments}
We thank Lionel Nowak, Carl van Vreeswijk {and Annette Witt} for valuable discussions and comments,
as well as the anonymous reviewers for helping us improving the manuscript.
DH thanks the Kavli Institute of Theoretical Physics at UC Santa Barbara and the program "Emerging Techniques in Neuroscience" for their warm hospitality while writing this manuscript.


\section*{Figure captions}

{\bf Figure 1.}
{\bf Schematic drawing of the model hypercolumns.} 
A: cartoon of the loop circuit among the 6 layers of striate cortex. 
Thalamo-recipient layers are indicated by pink shading. 
B: two-rings network, corresponding to a hypercolumn with 
interacting layers. LGN inputs are weaker toward the lower layer than toward 
the upper layer. 
C: the single ring network 
for each layer of the model hypercolumn. LGN inputs target both excitatory and 
inhibitory neurons. D: spatial profile of LGN input.
E: spatial modulation of the probability of 
connections between two cells in the same layer, separated by an angular 
distance $\Delta\vartheta$. Red line: excitatory connections. Blue line: 
inhibitory connections.  F: spatial modulation of the 
probability of connections between two cells in different layers, separated by 
an angular distance $\Delta\vartheta$. Red line: upper-to-lower 
layer excitatory connections and lower-to-upper excitatory connections 
toward excitatory neurons. Magenta line: lower-to-upper layer 
excitatory connections toward inhibitory neurons. Blue line: 
lower-to-upper and upper-to-lower layer inhibitory connections.  

{\bf Figure 2.}
{\bf Response tuning and contrast response.}  A: tuning curves for different contrast levels (re-centered average over $N_E = 4000$ 
excitatory neurons in upper layer). Solid lines represent Gaussian fits. B: contrast response functions (blue curve: average over $N_I = 1000$ inhibitory neurons in the upper layer; red curve: average over $N_E = 4000$ excitatory neurons in the upper layer). Solid lines represent hyperbolic ratio fits.

{\bf Figure 3.}
{\bf Low contrast dynamics.} Dynamics of the upper layer for the presentation 
of a 2\%-contrast stimulus.  
A: raster plot of the excitatory population activity and associated 
time-histogram of the rate of spiking cells. The histogram bar heights 
denote the fraction of upper layer excitatory cells that fire in the bin. Bin-size is 2 ms.
B: spike trains of 6 excitatory cells highly 
activated by the presented stimulus.  C: membrane potential traces for
two neurons stimulated simultaneously at close-to-preferred orientation
(2 top neurons of Panel B in red and green).  
D: pairwise correlations between spike trains 
(left, cyan histogram) and membrane potentials (right, blue histogram) of 
highly active neurons.

{\bf Figure 4.}
{\bf High contrast dynamics.} Dynamics of the upper layer for the presentation 
of a 95\%-contrast stimulus.  
A: raster plot of the excitatory population activity and associated 
time-histogram of the rate of spiking cells. The histogram bar heights 
denote the fraction of upper layer excitatory cells that fire in the bin. Bin-size is 2 ms.
B: spike trains of 6 excitatory cells highly 
activated by the presented stimulus.  C: membrane potential traces for
two neurons stimulated simultaneously at close-to-preferred orientation
(2 top neurons of Panel B in red and green).  
D: pairwise correlations
between spike trains (magenta histogram) and membrane potentials 
(red histogram) of highly active neurons. 

{\bf Figure 5.}
{\bf Pairwise crosscorrelations of spike trains and membrane potentials.} Autocorrelograms and pairwise crosscorrelograms of spiking activity and membrane potentials for three upper layer excitatory neurons. A: spiking activity, low contrast, $C= 2\%$. B: membrane potential, low contrast, $C= 2\%$. C: spiking activity, high contrast, $C= 95\%$. D: membrane potential, high contrast, $C= 95\%$. Auto- and crosscorrelograms are normalized (for zero time-lag, autocorrelograms peak at one and crosscorrelograms at the correlation coefficient). The units for the time-lag axis are ms. Colors are as in Figures~\ref{fig:LowContrast}D and~ \ref{fig:HighContrast}D. Rows and columns correspond to different neurons. The angular coordinates of the three neurons are $0^\circ$, $-10^\circ$ and $10^\circ$.

{\bf Figure 6.}
{\bf The Measure of synchrony as a function of the contrast and different
network size.}
A: The synchrony measure, $\chi$, increases abruptly with the stimulus contrast
$N = 10000$ (solid line) and $N = 40000$ dotted line).
B: The synchrony measure $\chi$ as a function of the network size
for spontaneous activity (zero contrast, grey line), low contrast 
(blue line) and high contrast (red line). 
The dashed line corresponds to a power-law decay with exponent -0.5.

{\bf Figure 7.}
{\bf The autocorrelograms of the local field potentials.} A--B: 
low contrast, $C= 2\%$. C--D: high contrast, $C=95\%$.  Scalings of non-normalized autocorrelograms are shown in B and D.
In both cases the damping of secondary peaks is faster for larger
network sizes. Zero-lag autocorrelation vanishes for large sizes at low contrast but not at high contrasts. Non-normalized autocorrelations are measured in $nA^2$. 

{\bf Figure 8.}
{\bf Spectral properties of the LFP and MUA for different contrasts.}
A: The power spectra are shown for the LFP induced by a stimulus at preferred 
orientation. Isolated peaks do not appear even for very high contrast stimuli.
B: The average coherence spectra are shown between the MUA and the LFP induced at a same location by a stimulus at preferred orientation. MUA-LFP coherence and LFP power are modulated by contrast changes in the same broad frequency range in the gamma band (30-100 Hz).  

{\bf Figure 9.}
{\bf Effects of the layer decoupling on the dynamics of the hypercolumn.} 
Changes for decreasing inter-layer coupling and for a stimulus at high 
contrast with preferred orientation.  A:  population average peak firing rate 
for the excitatory neurons in the upper layer.  B: synchrony level $\chi$.  
C: autocorrelograms of LFPs for intermediate strengths of the inter-layer 
coupling ($\Gamma = 0.8, 0.6$ and $0.2$).  D: corresponding LFP power spectra. 
E: autocorrelograms of LFP for preferred stimulation at high contrast for 
the case of fully uncoupled layers ($\Gamma = 0.0$).  F: corresponding LFP 
power spectrum. Spectra are also plotted for lower levels of contrast 
and are characterized by a narrow peak at a contrast-dependent frequency. 

{\bf Figure 10.}
{\bf Short-term response.} Population firing responses to repeated presentations of a high contrast stimulus for fully coupled layers (A--B, $\Gamma = 1$) and for fully uncoupled layers (C--D, $\Gamma = 0$).  A and C: peristimulus-time (PST) histograms, based on the firing responses of 500 cells to 1000 presentations of stimuli with optimal (or close to optimal) orientation.  B and D: examples of upper layer excitatory population responses for three presentations of the same stimulus.

{\bf Figure 11.}
{\bf Chaotic sensitivity to a single spike perturbation.} A black triangle denotes the time of a small perturbation to the network dynamics (for 95\% of contrast stimulus and fully-coupled layers, $\Gamma = 1$), in which a single spiking event is omitted. Already after the second oscillation cycle, the unperturbed and perturbed population dynamics have diverged, as visualized by the raster plot (A) and the population rate histogram (B) of the upper layer excitatory population. Blue color denotes unperturbed dynamics and red color perturbed dynamics.

\section*{Supporting information: captions}

\textsl{Supporting informations can be downloaded at the web address:}

\texttt{http://www.ploscompbiol.org/article/info\%3Adoi\%2F10.1371\%2Fjournal.pcbi.1002176}

\vspace{2em}

{\bf Figure S1.}
CV and firing rate distributions.  
Distributions of CVs and firing rates for highly active upper layer excitatory neurons (orientation preference within $\pm 5^\circ$ from stimulus) are here shown for the spontaneous activity ($C = 0\%$), for the low contrast regime ($C = 2\%$) and for the high contrast regime ($C = 95\%$). Distributions of CVs (A--C) and of firing rates (D--F), from top to bottom, for the spontaneous activity, for the low contrast regime and for the high contrast regime.  

\vspace{1em}
{\bf Figure S2.}
Heterogeneity of single neuron tuning curves. 
A: re-centered single neuron tuning curves for 3 upper layer excitatory neurons. B: distribution of peak rates.  C: distribution of tuning width.  D: distribution of tuning skewness. Distributions are relative to the upper layer excitatory population.

\vspace{1em}
{\bf Figure S3.}
Contrast invariance of tuning width.
Tuning curves normalized to peak height, for: fully coupled layers case (A), fully uncoupled layers case (B) and strong noise case (C; see Figures S13 and S14).  In cases A and B, contrast invariance is only approximate and does not hold for weak contrasts. Contrast invariance at low contrasts is improved in the strong noise case, in agreement with theory.

\vspace{1em}
{\bf Figure S4.}
Heterogeneity of single neuron contrast response functions. 
A: single neuron CRFs for 3 upper layer excitatory neurons.  B: distribution of saturation rates.  C: distribution of CRF steepness.  D: distribution of mid-range contrasts. Distributions are relative to the upper layer excitatory population.

\vspace{1em}
{\bf Figure S5.}
Correlation coefficients for spontaneous activity.
A: pairwise correlations (CCos) between membrane potentials. B: pairwise correlations between spike trains. CCos are computed as described in the Materials and Methods section of the main article.

\vspace{1em}
{\bf Figure S6.}
Oscillatory structure of induced LFPs. 
A--B: example LFP traces from our model, evoked by a stimulus with 2\% of contrast (A) or 95\% of contrast (B).
C--D: autocorrelograms and crosscorrelograms of evoked LFPs. For a fixed stimulus orientation, we monitor ACs and CCs of LFPs in regions responding preferentially to this stimulus orientation or to an orthogonal stimulus orientation. The analysis is performed for 2\% of contrast stimuli (C) or for 95\% of contrast stimuli (D). An oscillatory structure is present in LFP independently from spiking and is correlated over the entire ring.
E--F: LFP temporal decorrelation at intermediate contrast levels (E: $C=20\%$; F: $C=4\%$). Damping of secondary peaks is fast at any contrast. Units are $\mathrm{nA}^2$.

\vspace{1em}
{\bf Figure S7.}
Dynamics of the lower layer.
A: raster plots of the activity of the lower layer excitatory neurons (lower raster) and the upper layer excitatory neurons (upper raster) in the high contrast regime dynamics ($C = 95\%$). B: the latency between induced oscillations in the upper and in the lower layer is estimated through the crosscorrelogram of LFPs in the two layers (high contrast regime, $C = 95\%$). The upper layer advances the lower layer $\sim 2.8 \mathrm{\,\,ms}$ on average.

\vspace{1em}
{\bf Figure S8.} 
Induced responses for fully uncoupled layers: changes in tuning curves.
For $\Gamma = 0$ (full layer uncoupling), the dynamics of the upper layer is equivalent to the case where there is full inactivation of the lower layer.
After layer uncoupling, and consistently with reference [79] (Allison and Bonds, 1994), we observe changes in preferred orientation, peak response rates and tuning curve width and skewness of single neuron tuning curves. We report here distributions of parameter changes (vertical dotted lines denote average parameters for fully coupled layers, $\Gamma = 1$). A: distribution of preferred orientation shifts. Preferred orientation of individual cells can move clockwise or anti-clockwise within a range of $\sim\pm 30^\circ$ but the distribution of shifts is symmetric, with no significant change at the population level. B: distribution of peak firing rate changes. The mean peak rate change is weakly positive, reflecting the overall inhibitory nature of inter-layer coupling. C: distribution of broadness changes. On average, the width of tuning curves is slightly increased. D: distribution of skewness changes. Skewness changes are observed in both directions and their distribution is symmetric, with no significant change at the population level. In general, the large heterogeneity in the effects of layer uncoupling on tuning properties must be noted.

\vspace{1em}
{\bf Figure S9.} 
Induced responses for fully uncoupled layers: dynamical properties.
Response of the upper layer for the presentation 
of a 95\%-contrast stimulus.  A: raster plot of the excitatory population activity and associated 
time-histogram of the rate of spiking cells. The histogram bar heights 
denote the fraction of upper layer excitatory cells which firing in the bin. Bin-size is 2 ms.
B: spike trains of 6 excitatory cells highly 
activated by the presented stimulus.  C: membrane potential traces for
two neurons stimulated simultaneously at close-to-preferred orientation
(2 top neurons of Panel B in red and green). This dynamics is strongly synchronous and approximately periodic. For increasing network size, oscillations tend to become more periodic, and collective synchrony does not vanish (not shown).

\vspace{1em}
{\bf Figure S10.}
Numerical experiments for chaos assessment. 
All the methods are described in Text S1. A: estimation of the minimal embedding dimension. The fraction of false neighbors is plotted against the embedding dimension for a LFP time-series generated by a full contrast preferred orientation stimulus ($\Gamma =1$, $C=95\%$). $N=1000$ pairs of candidate neighbor points have been considered for each embedding dimension ($\epsilon < 10^{-9}$). A threshold of $R^* = 10^3$ has been taken. A single LFP time-series long 10 hours of real time, with a sampling rate of 0.01 ms has been used for the estimation. The resulting embedding dimension appears to be $m\ge 4$. B-C: extraction of the largest Lyapunov exponent $\lambda_{max}$ for the dynamics induced by a full contrast preferred orientation stimulus. The relative growth in time $\frac{\delta_t}{\delta_0}$ of the average separation between LFP trajectories originated from neighbor points is plotted against time, for various embedding dimensions (average over at least $N=1000$ pairs of neighbors per considered embedding dimension).  A section of exponentially fast growth (linear growth in a semilogarithmic plot, denoting deterministic chaos) is identified for sufficiently large embedding dimension in the case of a hypercolumn with interacting layers ($\Gamma = 1$, panel A), but not in the case of a hypercolumn without inter-layer interactions ($\Gamma = 0$, panel B).

\vspace{1em}
{\bf Figure S11.} 
Alternative parameter choices: network with increased symmetry. 
We assumed in the main text that the LGN input to the lower layer is weaker. We show here results for the case in which the LGN input rate to lower layer is the same as to the upper layer and in which the latency of the upper-to-lower layer connections is as short as the latency of lower-to-upper layer connections. A: raster plot of the evoked activity of the upper layer excitatory population for a 95\% level of contrast stimulus. B: autocorrelogram of the evoked LFP. Units are in $\mathrm{nA}^2$. Note that synchronous chaos is still present, as evidenced by the fast damping of LFP autocorrelogram. The lower and the upper layer have now the same average firing rate and are on average in an in-phase locking.

\vspace{1em}
{\bf Figure S12.} 
Alternative parameter choices: network with densified inhibition.  
We assumed in the main text that the probability of inhibitory connection is four times larger than the probability of excitatory connections. Some Experimental studies like reference [72], however, report a probability of inhibitory connection ten times larger than for excitatory connections. We show here results for a 1:10 ratio of excitatory to inhibitory connection probability (probabilities used are $p^{(0)} = 0.6$ and $p^{(1)} = 0.3$ for intra-layer inhibitory connections and $p^{(0)} = 0.3$ and $p^{(1)} = 0.0$ for inter-layer inhibitory connections). A: raster plot of the evoked activity of the upper layer excitatory population for a 95\% level of contrast stimulus. B: autocorrelogram of the evoked LFP (for different network sizes). Units are in $\mathrm{nA}^2$. Note that synchronous chaos is still present, as evidenced by the fast damping of LFP autocorrelogram, accelerating for larger network sizes. An additional effect of increased inhibitory density is a stronger tendency to resonate for low contrast stimuli. A weak ``hump'' at frequencies close to 45 Hz is visible even in the spectrum of spontaneous activity (not shown).

\vspace{1em}
{\bf Figure S13.} 
E-I connectivity is not required for the generation of oscillations. 
With our parameter choices, oscillations are generated thanks to delayed mutual inhibition. Excitatory neurons indeed are not required for the generation of oscillations, but are entrained by the oscillation paced by inhibitory cells. This can be proven by numerical simulations in which the activity of excitatory neurons is completely suppressed by a strong hyperpolarizing current (raster plot in panel A) or in which synapses from excitatory to inhibitory neurons are removed (raster plot in panel B). In both cases the drive to inhibitory neurons in order to maintain their rate of activity unchanged. Note that oscillations continue to exist and their frequency does not increase consistently.

\vspace{1em}
{\bf Figure S14.}
Fluctuations for different noise regimes. The fluctuation level of input currents can be controlled by acting on the input rates and peak synaptic conductances. Small peak coupling conductances and large input rates yield a quasi tonic input (``small''-variance noise). Conversely, stronger peak coupling conductances and smaller input rates give rise to input currents with similar average value but stronger fluctuations in time (``large''-variance noise). The net input conductance (red = excitatory, blue = inhibitory) of an upper layer excitatory neuron driven by a full contrast stimulus is shown in panels A (small-variance noise) and C (large-variance noise). Subthreshold voltage fluctuation strength is plotted against tuned LGN input rate in panels B (small-variance noise) and D (large-variance noise). The rate ranges are different for small- and large-variance noises, but are meant to correspond conventionally in both cases to the 0\%-100\% contrast range (see Tables S2 and S3). For both noise regimes, the mean excitatory conductance is of $\sim 3-5 \mathrm{nS}$ for the spontaneous activity and of $\sim 20 \mathrm{nS}$ for full contrast stimuli and the mean inhibitory conductance is of $\sim 4-6 \mathrm{nS}$ for the spontaneous activity and of $\sim 40-50 \mathrm{nS}$ for full contrast stimuli. At high contrast, however, fluctuations in conductance are much stronger for small-variance input noise, because recurrent inputs are highly synchronous. Sub-threshold voltage fluctuations at high contrast are comparably strong for both noise regimes, because for small-variance noise weaker fluctuations in the input are amplified by strong conductance fluctuations. At low contrast, when the dynamics is asynchronous for both noise regimes, voltage fluctuations are stronger for large-variance noise.

\vspace{1em}
{\bf Figure S15.}
High contrast dynamics for large-variance noise. Dynamics of the upper layer for the presentation of a 95\%-contrast stimulus. Input noise parameters are reported in Table S1. A: raster plot of the excitatory population activity and associated time-histogram of the rate of spiking cells. The histogram bar heights denote the fraction of upper layer excitatory cells firing in the bin. Bin-size is 2 ms. B: spike trains of 6 excitatory cells highly activated by the presented stimulus. C: membrane potential traces for two neurons stimulated simultaneously at close-to-preferred orientation (2 top neurons of Panel B in red and green). This dynamics is asynchronous, as indicated by the scaling analyses of Figure S16 C--D.

\vspace{1em}
{\bf Figure S16.}
Temporal decorrelation and spectra of LFPs for large-variance noise.  Input noise parameters are reported in Table S1. A: autocorrelogram of the LFP evoked by a high contrast stimulation. Units are $\mathrm{nA}^2$. B: power spectra of evoked LFP for various contrast levels. C: scaling with network size of the 95\%-contrast synchrony factor $\chi$. D: scaling with network size of the 95\%-contrast LFP autocorrelogram. The dashed line is a power-law with exponent -0.5. This scaling is indicative of  an asynchronous state. Units are $\mathrm{nA}^2$.

\vspace{1em}
{\bf Figure S17.} 
Alternative parameter choices: network with a non modulated spatial profile of inter-layer excitation. 
With the parameter choices assumed in the main text,  the integrated effect of the inter- layer coupling is inhibitory. It is however moderately excitatory between neurons in close vertical alignment, due to the strong spatial modulation of the inter- layer excitation profile. We show here results for the case in which the spatial modulation of inter-layer excitation is removed and an equivalent average level of inter-layer excitation is used, but spread across all the angular distances (i.e. $p^{(1)}_{\mbox{{\scriptsize inter-layer}}, E}= 0$). A: raster plot of the evoked activity of the upper layer excitatory population for a 95\% level of contrast stimulus. B: autocorrelogram of the evoked LFP. Note that synchronous chaos disappears, replaced by almost periodic oscillations, very similar to the case of uncoupled layers ($\Gamma = 0$, see Figure S9). Conversely, removal of inter-layer inhibition would further strengthen synchronized chaos (not shown).

\vspace{1em}
{\bf Text S1.} 
Detailed methods for chaos assessment. 
Section 1: Determination of the minimum embedding dimension. Section 2: Extraction of the largest Lyapunov exponent $\lambda_{max}$. 

\vspace{1em}
{\bf Text S2.}
Correspondence between contrast and LGN input rate. 
Extended description of the rationale behind the mapping between contrast and noise input parameters.

\vspace{1em}
{\bf Table S1.} Strong noise LGN input parameters.
Parameters of the LGN input to the network for the high contrast strong noise regime. See Table S3 for more details.

\vspace{1em}
{\bf Table S2.}  
Correspondence between $C$ and $R_0^{LGN}$ for small-variance noise. 
Correspondences are computed approximately, assuming that each cell receives 30 independent AMPA synaptic inputs from LGN (see Text S2). For the response of a single LGN cell we assumed $r_0 = 5$ Hz and $r_1 = 48$ Hz.

\vspace{1em}
{\bf Table S3.}   
Correspondence between $C$ and $R_0^{LGN}$ for large-variance noise. 
Correspondences are computed approximately, assuming that each cell receives 10 AMPA synapses from 3 independent LGN neurons (see Text S2). For the response of a single LGN cell we assumed $r_0 = 5$ Hz and $r_1 = 32$ Hz.

\newpage



\begin{figure}[!ht]
\begin{center}
\includegraphics[width=4in]{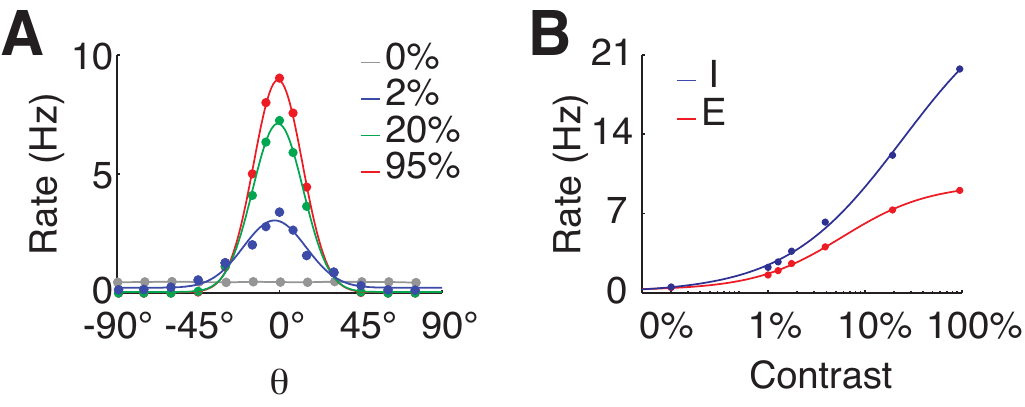}
\end{center}
\caption{}
\label{fig:TuningCRF}
\end{figure}


\begin{figure}[!ht]
\begin{center}
\includegraphics[width=4in]{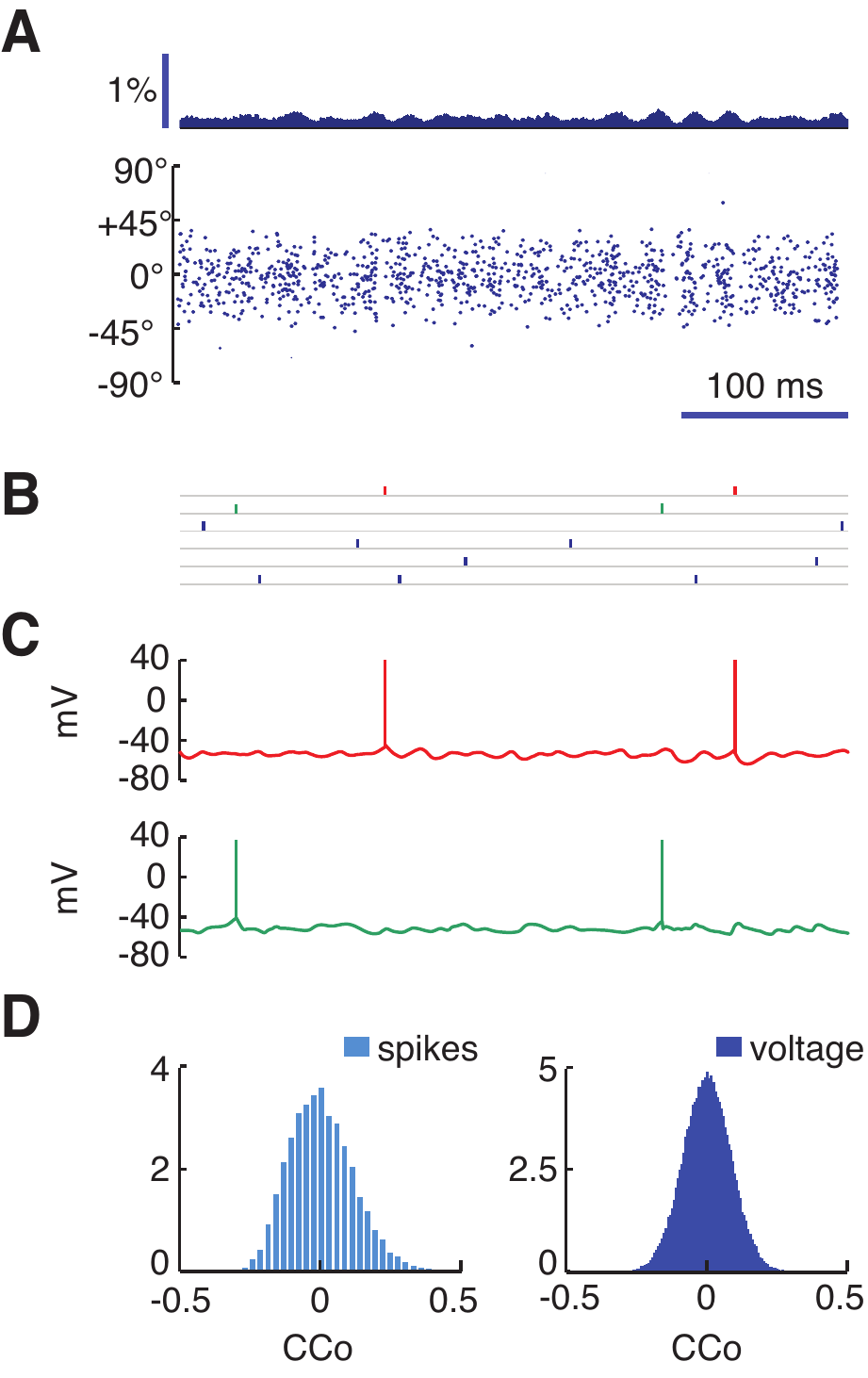}
\end{center}
\caption{}
\label{fig:LowContrast}
\end{figure}


\begin{figure}[!ht]
\begin{center}
\includegraphics[width=4in]{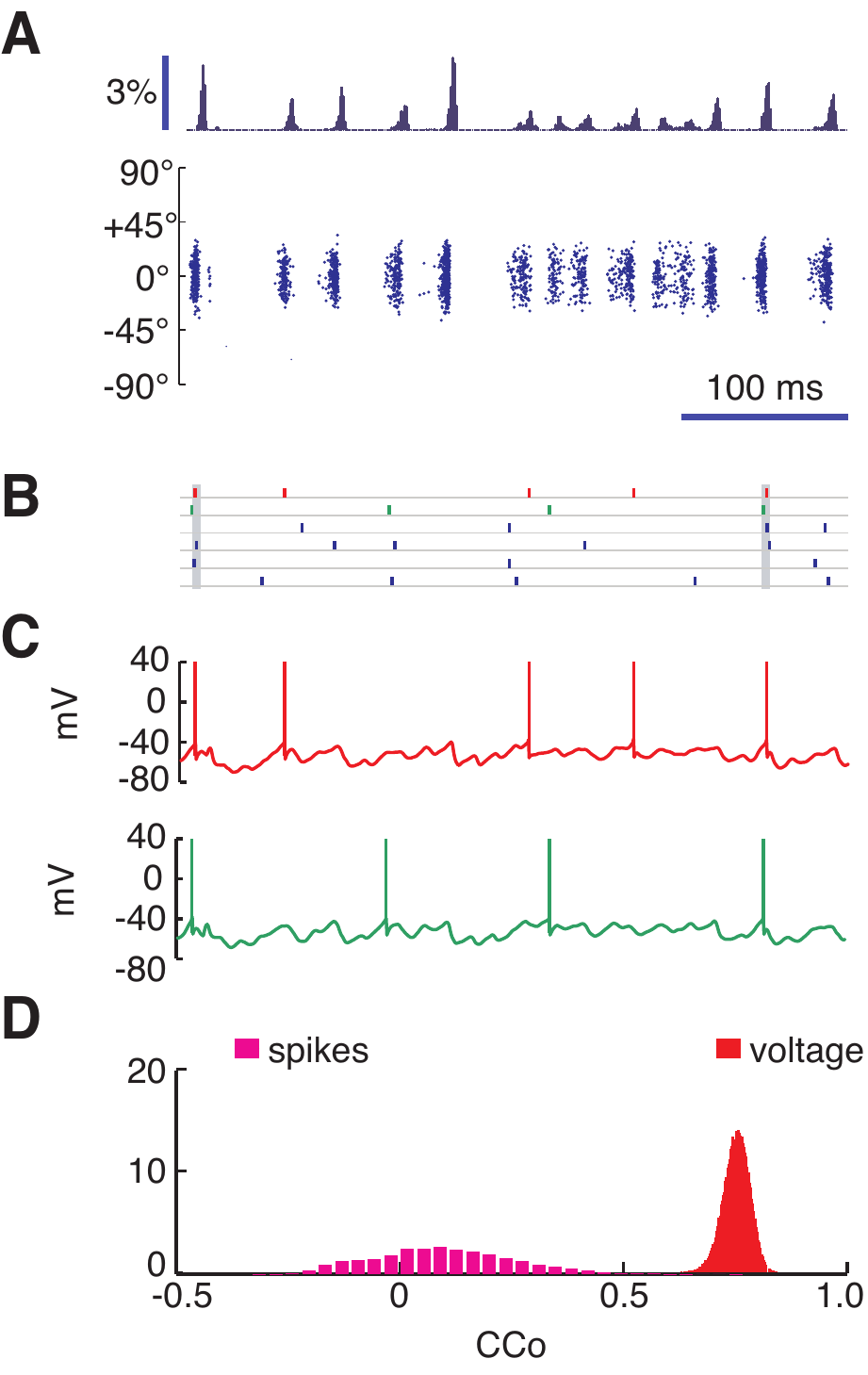}
\end{center}
\caption{}
\label{fig:HighContrast}
\end{figure}


\begin{figure}[!ht]
\begin{center}
\includegraphics[width=4in]{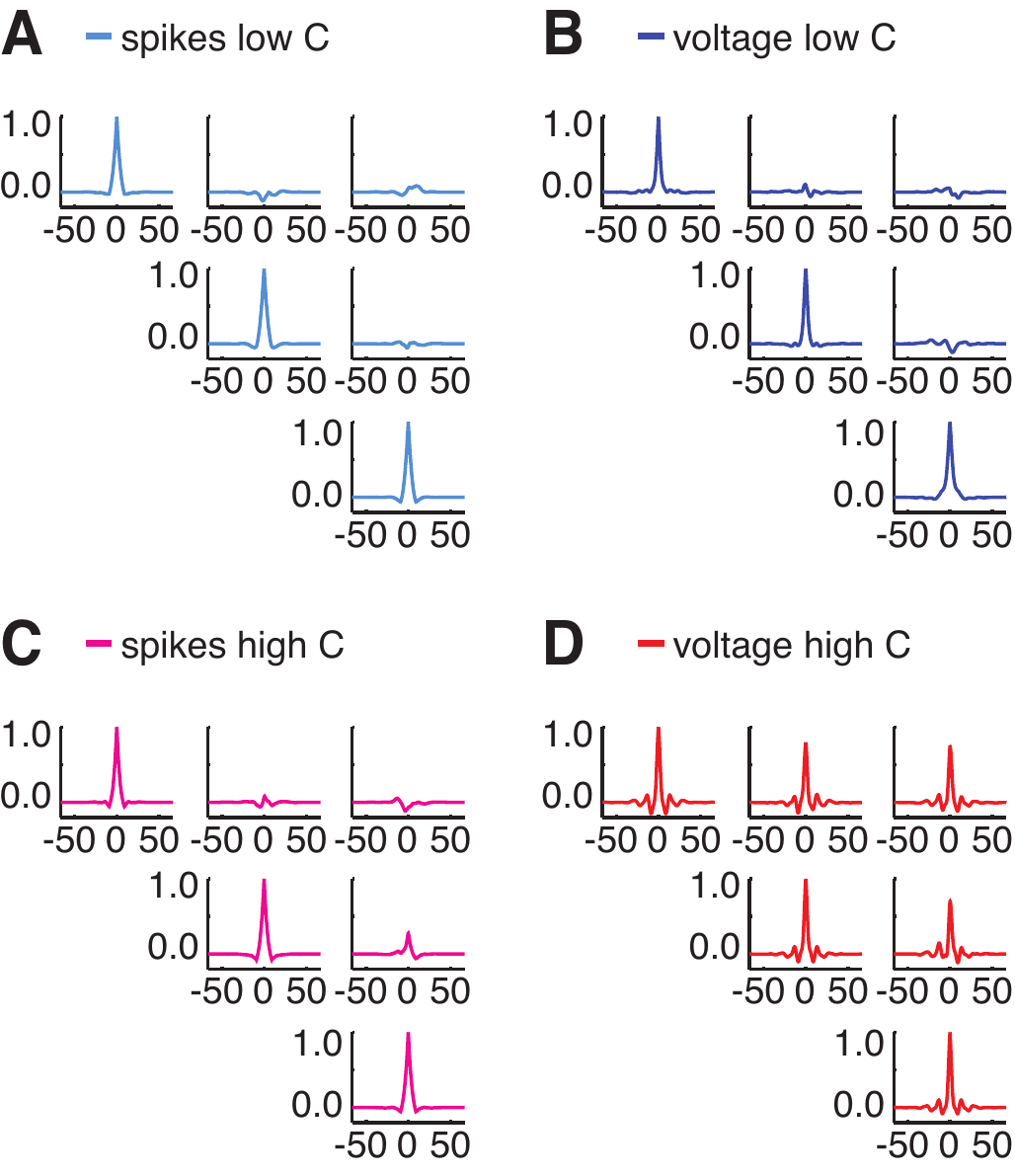}
\end{center}
\caption{}
\label{fig:PairwiseCCs}
\end{figure}


\begin{figure}[!ht]
\begin{center}
\includegraphics[width=4in]{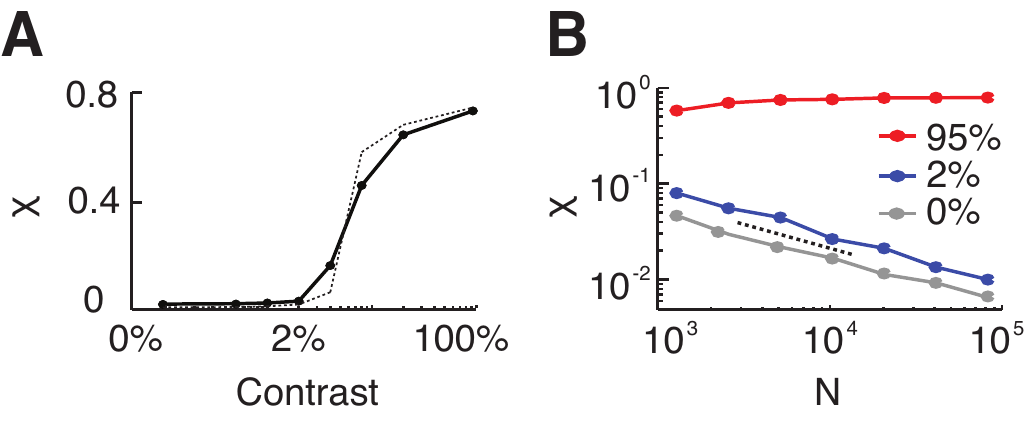}
\end{center}
\caption{}
\label{fig:Chi}
\end{figure}


\begin{figure}[!ht]
\begin{center}
\includegraphics[width=4in]{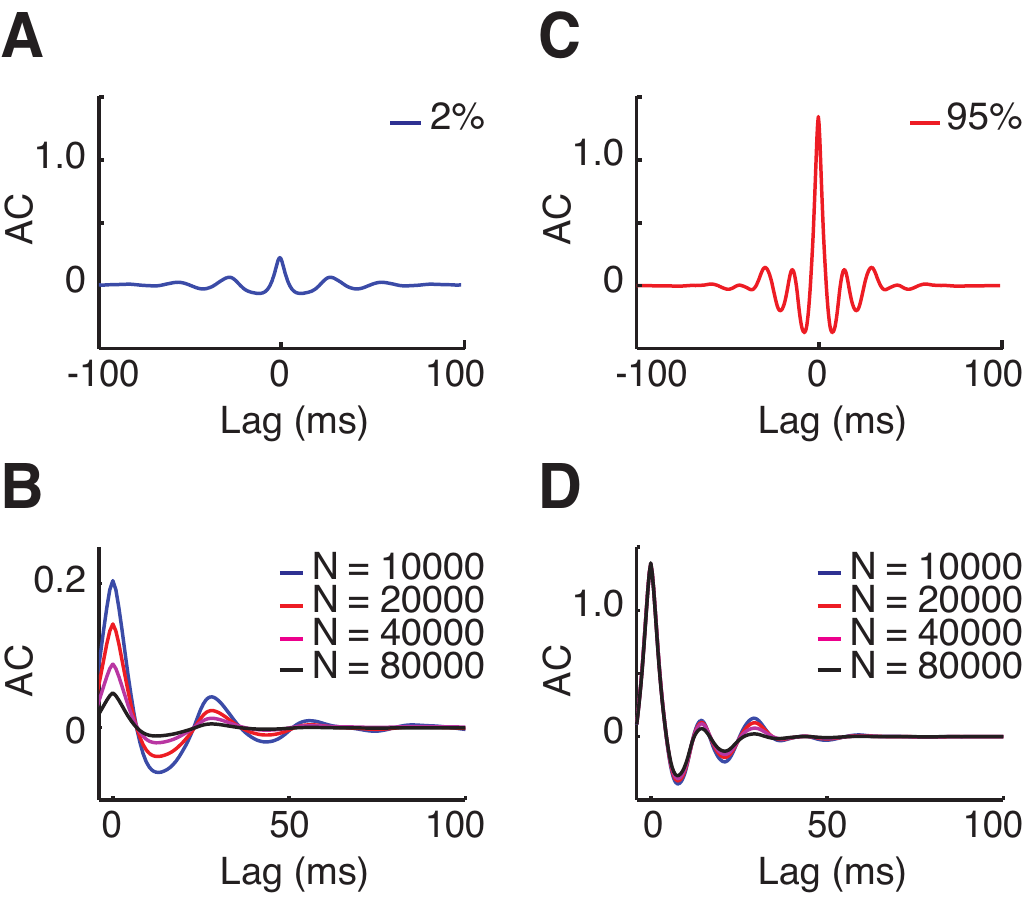}
\end{center}
\caption{}
\label{fig:LFPAC}
\end{figure}


\begin{figure}[!ht]
\begin{center}
\includegraphics[width=4in]{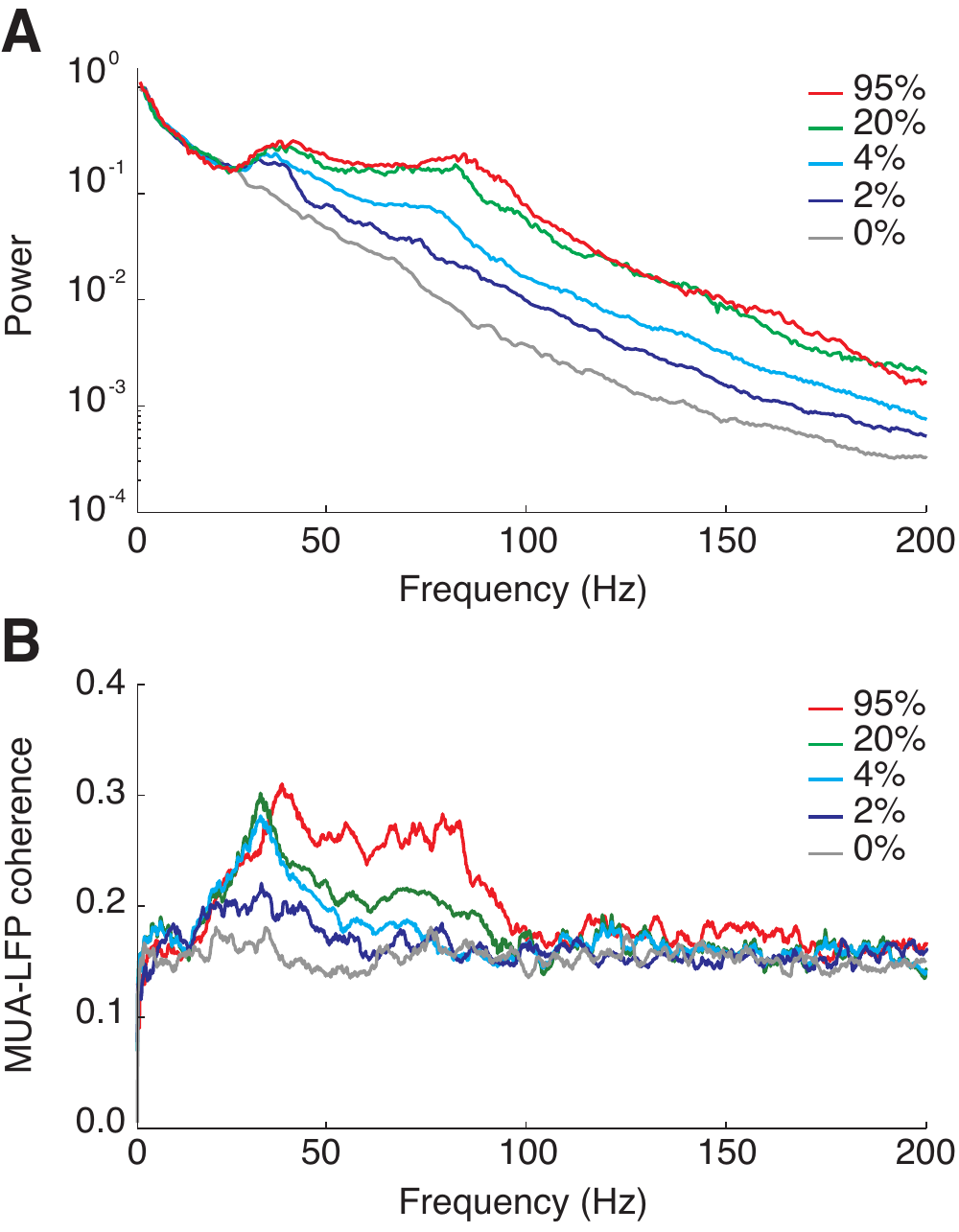}
\end{center}
\caption{}
\label{fig:LFP}
\end{figure}


\begin{figure}[!ht]
\begin{center}
\includegraphics[width=4in]{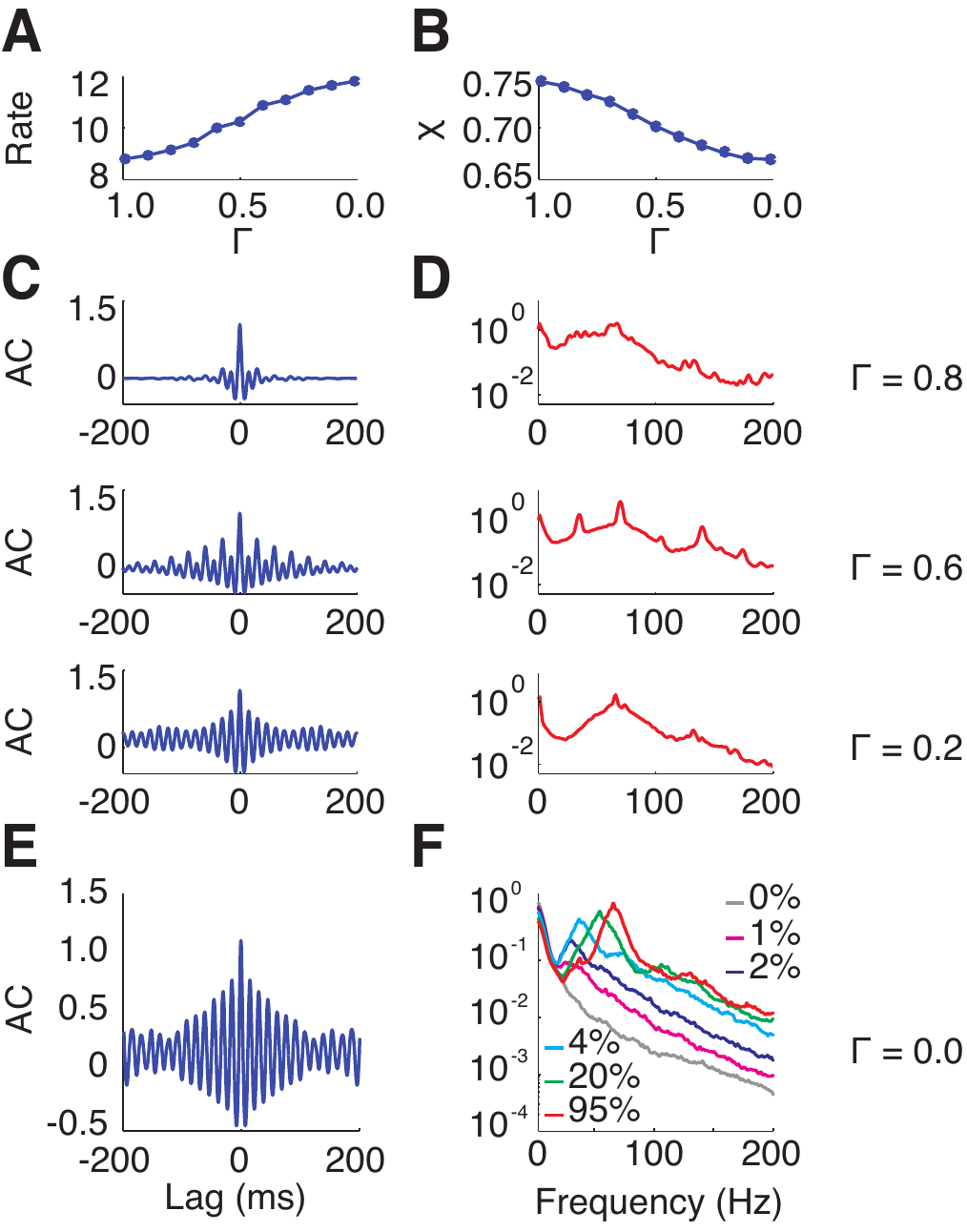}
\end{center}
\caption{}
\label{fig:Decoupling}
\end{figure}


\begin{figure}[!ht]
\begin{center}
\includegraphics[width=4in]{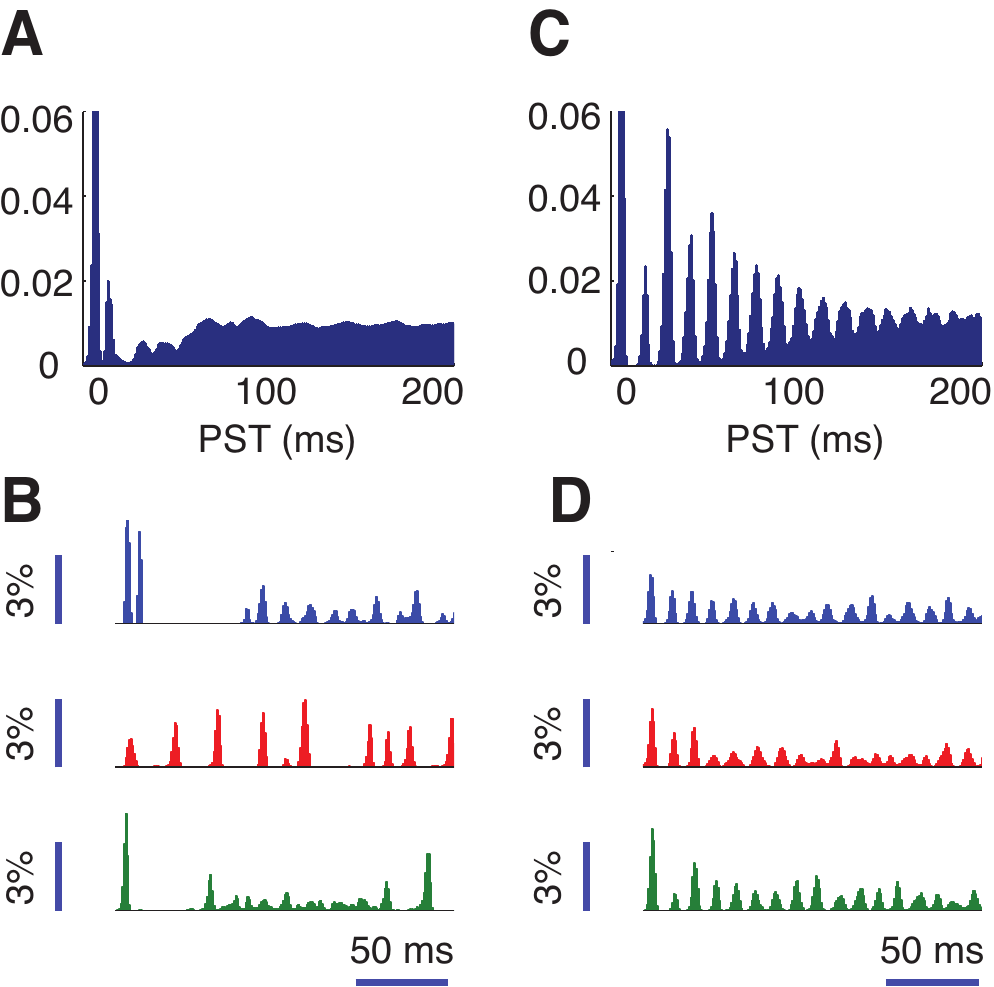}
\end{center}
\caption{}
\label{fig:PSTH}
\end{figure}


\begin{figure}[!ht]
\begin{center}
\includegraphics[width=4in]{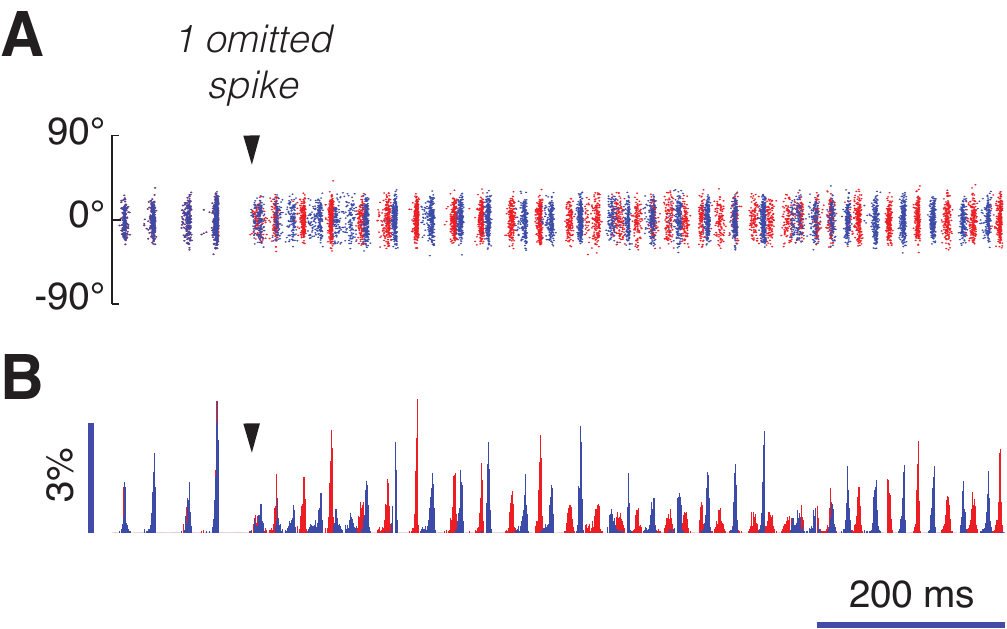}
\end{center}
\caption{}
\label{fig:Chaos}
\end{figure}

\section*{Tables}

\begin{table}[!ht]
\caption{
\bf{Parameters for model neurons}}
\begin{tabular}{|c|c|c|}\hline
\, 			& 	Excitatory & 	Inhibitory \\
\hline\hline
$\tau_m^0$ 	& 	23.30 ms	& 	11.65 ms	\\
\hline
$C$ 			& 	0.26 nF	&	0.13 nF 	\\
\hline
$V_L^0$		&	-57.8 mV	&	-57.8 mV	\\
\hline
$V_T^0$		&	-45.2 mV	&	-45.2 mV	\\
\hline
$\Delta T$ 	&	1.2 ms	&	1.2 ms	\\
\hline
\end{tabular}   
\begin{flushleft}
Parameters (without time dependency) of model excitatory and inhibitory EIF neurons.
\end{flushleft}
\label{tab:SingleNeuronStatic}
\end{table}

\begin{table}[!ht]
\caption{
\bf{Soft refractoriness parameters}}
\begin{tabular}{|c|c|c|c|c|}\hline
$x$			& 	$A_x$ 	& 	$\tau_{A,x}$ & 	$B_x$ 	& 	$\tau_{B,x}$ \\
\hline\hline
$V_L$		&	22.9 mV	&	14.7 ms	&	13.5 mV	&	76.2 ms	\\
\hline
$V_T$		&	10.0 mV	&	17.7 ms	&	---		&	---	\\
\hline
$1 / \tau_m$ 	& 	0.14 ms$^{-1}$		& 	14.3 ms	& ---	&	---	\\
\hline
\end{tabular}   
\begin{flushleft}
Parameters of time-dependent after-spike relaxation of excitatory EIF model neurons
\end{flushleft}
\label{tab:SingleNeuronDynamic}
\end{table}

\begin{table}[!ht]
\caption{
\bf{Synaptic time-constants and efficacies}}
\begin{tabular}{|c|c|c|c|c|}\hline
\, 		& 	$\tau_r$ (ms) 	& $\tau_d$ (ms) & g (nS) &PSP (mV)\\
\hline\hline
AMPA on excitatory 	& 	1 	& 	3 	& 	1.0 	&	0.84	\\
\hline
AMPA on  inhibitory	 & 	1 	& 	3 	& 	1.5 	&	2.07	\\
\hline
GABA on excitatory	& 	1 	& 	4 	& 	4.0	&	1.13	\\
\hline
GABA on inhibitory 	& 	1 	& 	2 	& 	4.0	&	1.36	\\
\hline
NMDA on excitatory & 	3 	& 	80 	& 	0.14	&	0.50	\\
\hline
\end{tabular}   
\begin{flushleft}
Synaptic parameters for a network of $N_E = 4000$ neurons and $N_I = 1000$ neurons: synaptic rise ($\tau_r$) and decay ($\tau_d$) times, peak synaptic conductance ($g$) and peak postsynaptic potential $PSP$.
\end{flushleft}
\label{tab:Condu}
\end{table}

\begin{table}[!ht]
\caption{
\bf{Synaptic latencies}}
\begin{tabular}{|c|c|}\hline
\, 									& 	$d$ (ms)\\
\hline\hline
Intra-layer synapse 						& 	1.0	\\
\hline
Inter-layer synapse (upper to lower layer) 	& 	3.0	\\
\hline
Inter-layer synapse (lower to upper layer)		& 	1.0	\\
\hline
\end{tabular}   
\begin{flushleft}
Synaptic latencies ($d$) depending on the relative position of pre- and post-synaptic neurons.
\end{flushleft}
\label{tab:Latencies}
\end{table}

\begin{table}[!ht]
\caption{
\bf{Probabilities of connection}}
\begin{tabular}{|c|c|c|c|c|}
\hline
\, & $p^{(0)}$ & $p^{(1)}$ & Mean target E cells & Mean target I cells\\
\hline\hline
Intra-layer E to E or I & 0.06 & 0.06 & $\sim$240 & $\sim$60\\
\hline
Intra-layer I to E or I & 0.24 & 0.12 & $\sim$960 & $\sim$240\\
\hline
Upper E to Lower E or I & 0.06 & 0.18 & $\sim$240 & $\sim$60\\
\hline
Lower  E to Upper E only& 0.06 & 0.18 & $\sim$240 & ---\\
\hline
Lower  E to Upper I only& 0.07 &  0.16 & --- & $\sim$70\\
\hline
Inter-layer I to E or I & 0.12 & 0.00 & $\sim$480 & $\sim$120\\ 
\hline
\end{tabular}   
\begin{flushleft}
Probabilities of connection. The connection probability parameters $p^{(0)}$ and $p^{(1)}$ are given for a network size of $N_E = 4000$ and $N_I = 1000$ per layer. 
\end{flushleft}
\label{tab:ProbConn}
\end{table}

\begin{table}[!ht]
\caption{
\bf{LGN input}}
\begin{tabular}{|c|c|}\hline
$g_{LGN}$ 	& 	1 nS		\\
\hline
$R_0$ 		& 	150 Hz	\\
\hline
$R_1$		& 	2850 Hz	\\
\hline
$r_0$		& 	5 Hz		\\
\hline
$r_1$		&	48 Hz	\\
\hline
\end{tabular}   
\begin{flushleft}
Parameters of the LGN input to the network $R_{LGN}(C)$. See Text and Tables S10 for more details.
\end{flushleft}
\label{tab:Contrast}
\end{table}

\begin{table}[!ht]
\caption{
\bf{Background cortical noise}}
\begin{tabular}{|c|c|}\hline
$g_{bg}$		& 		10 nS	\\
\hline
$\mu_{bg}$	& 		10 Hz	\\
\hline
$\sigma_{bg}$	& 		1 Hz		\\
\hline
$\tau_{bg}$	&		10 ms	\\
\hline
\end{tabular}   
\begin{flushleft}
Parameters of the background cortical input $R_{bg}$.
\end{flushleft}
\label{tab:Background}
\end{table}

\end{document}